# Target and Conditional Sensitivity Analysis with Emphasis on Dependence Measures

Hugo Raguet[1] and Amandine Marrel[1]

**Abstract**

In the context of sensitivity analysis of complex phenomena in presence of uncertainty, we motivate and precise the idea of orienting the analysis towards a critical domain of the studied phenomenon. We make a brief history of related approaches in the literature, and propose a more general and systematic approach. Nonparametric measures of dependence being well-suited to this approach, we also make a review of available methods and of their use for sensitivity analysis, and clarify some of their properties. As a byproduct, we notably describe a new way of computing correlation ratios for Sobol' indices, which does not require specific experience plans nor rely on independence of the input factors. Finally, we show on synthetic numerical experiments both the interest of target and conditional sensitivity analysis, and the relevance of the dependence measures.

## 1 Introduction

The variety of approaches and applications of sensitivity analysis brings forth a diversity of objects and terms. Through brief literature review, we try to clarify here the goals and extent of our work.

In the classical framework, we assume the modeling of a phenomenon $Y$ depending on a set of *factors* $(X_i)_{1 \leq i \leq d}$ following a deterministic relation $Y \stackrel{\text{def}}{=} f(X_1, \ldots, X_d)$. Uncertainties are taken into account by modeling the factors as random variables, defined over the same implicit probability space $(\Omega, \mathfrak{F}, P)$. It will also be convenient to consider a generic random variable $X$, usually standing for a group of one or several factors. For such a variable, we note its range $\mathscr{X} \stackrel{\text{def}}{=} \text{ran}(X)$ and its law $P_X \stackrel{\text{def}}{=} P \circ X^{-1}$ is the (Borel) probability measure that it induces over $\mathscr{X}$. We also particularize $\mathscr{Y} \stackrel{\text{def}}{=} \text{ran}(Y)$.

### 1.1 Object and Goals

*Global sensitivity analysis* aims at measuring how the variations of one or several factors contribute to the variation of the studied phenomenon, over the whole domain of possible values of the factors and of the phenomenon. In contrast, *target sensitivity analysis*, as we define it, still considers the entire domain of the factors, but aims at measuring their influence over a restricted domain of the studied phenomenon, and in particular over the *occurrence* of the phenomenon in this restricted domain. Such domain of interest would usually be extreme and relatively rare, constituting a risk or an opportunity; we call it *critical domain*, noted $\mathscr{C} \subset \mathscr{Y}$ and associated to a *critical probability* $P(Y \in \mathscr{C}) = P_Y(\mathscr{C})$. Alternatively, *conditional sensitivity analysis* evaluates the influence of the factors *within* the critical domain only, ignoring what happens outside. Let us underline that those two notions can widely differ.

It seems to us that there are numerous, direct applications, especially for, but not restricted to, industrial safety. Still, while global sensitivity analysis has been an active research field for several decades, it seems that target sensitivity analysis is less understood, and until recently has not been studied systematically as such. This is why we introduce our own terminology, which we discuss along with the description of similar concepts that we identify in the literature.

[1]CEA, DEN, DER, F-13108, Saint-Paul-lez-Durance, France; hugo.raguet@gmail.com

Many authors agree with Saltelli et al. (2008) to distinguish several use of sensitivity analysis. First, the *ranking* of factors by importance is the starting point of any application. Identifying the factors which are most influential in a phenomenon might help understanding it or guide resource investment for controlling it. Then, *screening* the factors for insignificant ones is considered, for instance for model simplification. This sometimes calls on statistical tests, which might be practical but cannot be satisfying for all applications. In our experience, screening is often in practice interpretation of ranking, either through the expertise of the practitioner or with some cross-validation process. Finally, *factor mapping* is often described as a finer identification of functional relationship between the specific domains of values of the factors and of the phenomenon. This is reminiscent of target sensitivity analysis, since certain domains of values of the phenomenon are particularized. However, determining which factors contribute most in the occurrence of the phenomenon in a given domain is a different task than determining *which values* of these factors are responsible of such occurrence. In general, target sensitivity analysis only deals with the former.

Finally, let us point out that we are mostly interested in phenomena influenced by many factors and of which only limited understanding is available. Typical situations include complex systems observed through heavy computer simulations or costly physical measures.

## 1.2 Review on Existing Approaches

We propose here a coarse classification of methods relating to target sensitivity analysis, according to both chronological and methodological criteria.

### 1.2.1 Regional Sensitivity Analysis

The very notion of target sensitivity analysis dates back at least to Spear and Hornberger (1980), motivated by environmental science applications. The proposed methodology compares the distribution of the factors within the critical domain against their distribution outside. The authors choose to use the *Kolmogorov distance*, almost systematically reused ever since:

$$\sup_{x \in \mathcal{X}} \left| F_{X|Y \in \mathcal{C}}(x) - F_{X|Y \in \mathcal{Y} \setminus \mathcal{C}}(x) \right|,$$

where $F_{X|A}$ is the *cumulative distribution function* of a real random variable $X$ (i.e. $\mathcal{X} \subseteq \mathbb{R}$) conditioned by an event $A \in \mathfrak{F}$ of nonzero probability (see § 4.1.2 for details).

They call it *regional sensitivity analysis*, or sometimes generalized sensitivity analysis. The former name could fit our purpose, if it was not for two inconveniences. First, it evokes more of a sensitivity analysis within the critical domain (what we call conditional sensitivity analysis) rather than its occurrence; second, for the past three decades in the literature it referred exclusively to the above methodology. It appears to be the generalization of no other method, explaining why the alternative name is not used anymore. Finally, one may encounter the term *Monte Carlo Filtering*, which might be vague and restrictive.

Comparing distributions conditionally to the critical domain seems a good choice for target sensitivity analysis. It involves only two conditionings, which facilitates its estimation, for instance with Monte Carlo method. One difficulty, mentioned by the authors and common to all target sensitivity methods, arises when the critical probability is low. Another deficiency pointed out by the authors is the difficulty to study factors in interaction. From this viewpoint, observe that a metric comparing cumulative distribution functions can be extended to multidimensional settings, which would allow to regroup several factors. However, the particular metric used here, namely the supremum norm over the differences, is sensitive to outliers. Both aspects make it particularly unsuitable for categorical factors.

Strangely enough, regional sensitivity is mostly used in the literature as a mean of global sensitivity analysis, the partition of the domain of values of the phenomenon into several regions losing its original sense and becoming more or less arbitrary.

### 1.2.2 Reliability Sensitivity Analysis

Another field dealing with target sensitivity analysis is motivated by applications in structural reliability, where the term *reliability sensitivity analysis* is commonly used. In this context, critical domains are failure domains, and the developed methods are influenced by two typical features: failure probabilities are small in comparison to the number of available observations, and the probability distributions of the factors are assumed to be known.

The first methods developed seek, in a suitable transformation of the factors space, to determine a "most probable failure point", and to estimate the critical probability from linear or quadratic approximations of the boundary of the critical domain around that point. This yields the *first-* and *second-order reliability methods*, reviewed by Rackwitz (2001). It is possible to give to each factor an importance measure based on the position of the most probable failure point. The geometrical assumption about the failure domain seems however restrictive, implying in particular that the factors have a monotonous effect on the phenomenon.

The sensitivity measures which later prevail in the field are based on derivatives of the critical probability, with respect to the parameters defining the probability laws of the factors or of their transformation. This framework seems once again restrictive for our purpose. Nonetheless, the approaches developed in parallel for dealing with low critical probabilities deserves to be incidentally noted, because they could be adapted to other sensitivity measures. Let us mention the methods based on *importance sampling* (see for instance the adaptation of Wu, 1994), as well as the approach of *sequential Monte Carlo* as proposed by Au and Beck (2001), who call it *subset simulation*. As further developed by Song et al. (2009) and Cérou et al. (2012), the latter is based on Markov chain Monte Carlo with the Metropolis–Hasting algorithm.

Still in the reliability context, the Ph.D. dissertation of Lemaître (2014) is the first systematic study of target sensitivity analysis. With this purpose in mind, the author compare more general methods of global sensitivity analysis, of which we give a brief overview below. We can already mention that he identifies the need to transform the variable modeling the phenomenon into a binary variable encoding the occurrence in the critical domain; that is $1_{\mathscr{C}}(Y)$, where $1_{\mathscr{C}}: y \mapsto 1$ if $y \in \mathscr{C}$, 0 otherwise. This is one of the approach on which we focus in this work (see § 4.1).

A first sensitivity analysis method considered is the estimation of (square) *correlation ratio* between the factors and the phenomenon (real, with finite variance),

$$\eta^2(X, Y) \stackrel{\text{def}}{=} \frac{\mathrm{V}(\mathrm{E}[Y \mid X])}{\mathrm{V}(Y)} \, . \tag{1.1}$$

Resulting quantities are often called Sobol' (1990, 1993) indices. These are nowadays standard for global sensitivity analysis, notably because they can be interpreted in terms of decomposition of the variance of the studied phenomenon. Lemaître shows how these indices applied to the binary transformation of the observed phenomenon, $\eta^2(X, 1_{\mathscr{C}}(Y))$, are relevant at least for cases that are simple and where the number of available observations is high enough.

A second method is based on the *total variation* which we develop later (see § 3.4), once again applied to the binary transformation. Unfortunately, the proposed estimation methods might be inadequate and the analysis is too brief; the author mentions a "positive bias" without further explanations.

Another set of methods is based on *binary classification trees*. The author lists many ways of defining classification trees, and even more ways of deducing sensitivity indices. This indicates a lack of generality and robustness, actually revealed by some numerical experiments. For the sake of brevity, we do not elaborate here and invite the interested reader to refer to the dissertation for more details.

Then, Lemaître takes over regional sensitivity described above with some modifications. He compares the probability laws conditionally to the critical domain against the (known) marginal probability laws. In addition to Kolmogorov distance, he tries other discrepancy measures between cumulative distribution functions classically used in statistical tests, namely Cramér–Von Mises and Anderson–Darling, and shows that this choice can influence the importance ranking of factors. More importantly, he suggests that sequential Monte Carlo approaches are well adapted to methods based on comparisons of factors distributions conditionally to the critical domain.

Finally, closer to the classical sensitivity measures for reliability mentioned above, the author proposes its own measures, quantifying how modifications of the factors probability laws impact on the critical probability; published elsewhere by Lemaître et al. (2015). Actually, this approach could be used to study any other statistics than the critical probability. It is relevent in a framework where one wants to quantify uncertainties due to estimation errors on the model's parameters. In our framework, where we do not even assume the knowledge of the factors probability laws, such perturbations of the laws seem somewhat artificial.

Altogether, this Ph.D. dissertation is an interesting entry point to target sensitivity analysis. However, more numerical experiments seem necessary in order to conclude about the advantages and drawbacks of the different considered approaches, and those which should be retained for further improvements and comparisons are not clearly identified.

### 1.2.3 Sensitivity Analysis of a Specific Statistic

Another recent approach for target sensitivity analysis is due to Fort et al. (2016). Their formulation is more precise than ours: they are interested in the sensitivity of *an estimator of a statistical quantity* of the studied phenomenon. For this, they introduce the term of *goal-oriented sensibility analysis*.[1]

From the relations $V(E[Y|X]) = E\left((E[Y|X] - E[Y])^2\right) = V(Y) - E(V[Y|X])$, the authors show how the correlation ratio, equation 1.1, is, in their sense, a measure adapted to the sensitivity of the *expectation* of the phenomenon: indeed, it measures a distance between expectations, quantified by a difference of variances. Now for a generic real random variable $Y$, expectation and variance can be defined through an optimization problem, $E[Y] = \text{argmin}_{\theta \in \mathbb{R}} E((Y-\theta)^2)$ and $V[Y] = \min_{\theta \in \mathbb{R}} E((Y-\theta)^2)$, where the functional $(y, \theta) \mapsto (y-\theta)^2$ plays the role of a *contrast function*.

The generalization of the correlation ratio to a statistic defined by another contrast function $\psi$ becomes[2] $\min_{\theta \in \mathbb{R}} E(\psi(Y,\theta)) - E(\min_{\theta \in \mathbb{R}} E[\psi(Y,\theta)|X])$. In practice, in order to study extreme values, they focus on the *quantiles* of the phenomenon, considering for a level $\alpha \in ]0,1[$, the contrast function $(y,\theta) \mapsto (y-\theta)(1_{\{y \leq \theta\}} - \alpha)$. However, resulting indices turn out to be difficult to estimate, as shown by the recent developments of Browne et al. (2017) and Maume-Deschamps and Niang (2018).

---

[1] Beware that this term already exists in the literature referring to tools of different nature.
[2] Provided that the random variable $\min_{\theta \in \mathbb{R}} E[\psi(Y,\theta)|X]$ is well defined.

Les us mention that Kucherenko and Song (2016) propose another adaptation of the correlation ratio to analysis of sensitivity of quantiles, more direct: expectations are simply replaced by quantiles[3] of level $\alpha \in\, ]0,1[$, $\mathrm{E}\!\left(\left(F_{Y|X}^{-1}(\alpha) - F_Y^{-1}(\alpha)\right)^2\right)$, where $F^{-1}$ is the generalized inverse of a cumulative distribution function. As its estimation is also difficult, the authors propose to approximate the quantiles conditionally to factors values by a rough form of kernel method.

At last, let us add that the quantile is a peculiar notion and in our opinion, its use for sensitivity analysis raises some troubles. Beyond difficulty of definition and estimation, these tools are adapted only to phenomena which are unidimensional and continuous. Moreover, the "sensitivity of a quantile" has a less straightforward interpretation than the sensitivity of the occurrence of a phenomenon, or of the variation of a phenomenon, in a critical domain.

### 1.3  Contributions and Outline of the Paper

The majority of the methods previously described are originally developed for particular applications; we would like to make abstraction of the problem to get more general methods. To this end, rather than defining or enhancing specific methods, we seek modifications or generalizations of global sensitivity analysis tools, which would be adapted to target or conditional sensitivity analysis.

Such modifications boil down, for a given analysis tool considered, to weighting the observations according to the critical domain. The weights can operate following two principles: either as a *transformation* of the phenomenon prior to the application of the tool, or as a *modification* of the parameters and objects which define the tool itself. This includes, but is not restricted to, the natural notion of conditioning. Several variations around these principles are presented along § 4; before that, the actual analysis tools must be introduced.

We describe in § 2 the sensitivity indices based on correlation ratio, the popular *Sobol' indices*. Now, it appears that sensitivity analysis based on *nonparametric dependence measures*, recently advocated by Da Veiga (2015), is particularly adapted to our framework. This will retain most our attention in the following, starting from § 3, where we review the available methods for measuring statistical dependence and detail some of their use in the context of sensitivity analysis.

Practical estimation is also one of our concern, and is addressed in parallel to the development of our sensitivity measures. However, estimation of dependence measures is an active area of research, into which we do not delve in this work; in particular, we do not discuss convergence properties. As for computational cost, we recall here that in our context, the cost of data acquisition is considered the most limiting factor: the number of available observations is typically in the order of hundreds to thousands. In consequence, computational load of the analysis tools themselves is not regarded as crucial.

Finally, we give along § 5 numerical evaluations of our methods on various synthetic data.

Naturally, we do not pretend to exhaustiveness; even within the adopted methodology described above, since we cannot evaluate in this work all existing dependence measures (see our review § 3.1). In addition, other popular approaches of global sensitivity analysis could be adapted to target or conditional sensitivity analysis. We voluntarily set those aside for brevity, but we would like to make here a few comments about two approaches described above, which seems promising to us and ought to be more deeply studied in future works.

---

[3]The random variable $F_{Y|X}^{-1}$ is now defined through conditional distribution.

The first is the core idea of regional sensitivity analysis presented in § 1.2.1. To us, the most practical aspect is that the involved distributions are conditioned by only two events, $\{Y \in \mathscr{C}\}$ and its complementary, the probability of the former being potentially low, but never negligible since $\mathscr{C}$ is a region of interest. Besides possible generalizations of the Kolmogorov distance, one should consider other measures of discrepancy between probability distributions, such as the ones used in our study and described along § 3. Moreover, as suggested by Lemaître (2014), its potential estimation through sequential Monte Carlo algorithm is a valuable advantage for very low critical probabilities.

Second, note that $P_Y(\mathscr{C})$ is a statistic on $Y$ which can be derived from the contrast function $(y, \theta) \mapsto (1_{\mathscr{C}}(y) - \theta)^2$. Adapting the method of Fort et al. (2016) and skipping some derivations, we get the sensitivity measure $P_Y(\mathscr{C})(1 - P_Y(\mathscr{C})) - E(P_{Y|X}(\mathscr{C})(1 - P_{Y|X}(\mathscr{C})))$; this is (proportional to) the correlation ratio $\eta^2(X, 1_{\mathscr{C}}(Y))$, which is not surprising since the critical probability is nothing but the expectation of the binary transformation. Just as Kucherenko and Song (2016) do for the quantiles, one can also directly compare the critical and conditional probabilities, for instance with $E(|P_Y(\mathscr{C}) - P_{Y|X}(\mathscr{C})|)$; actually, this can be shown to be (proportional to) the dependence measure based on the total variation presented below, applied to the binary transformation $1_{\mathscr{C}}(Y)$. We think that yet new methods could be derived from the above approaches; but it should be kept in mind that conditioning by the values of the factors raises estimation problems, especially for low critical probabilities.

## 2 Sensitivity Analysis with Correlation Ratio

Given a group of factors $I \subset \{1, \ldots, d\}$, we write $X_I \stackrel{\text{def}}{=} (X_i)_{i \in I}$ for the corresponding random tuple, and $^cI \stackrel{\text{def}}{=} \{1, \ldots, d\} \setminus I$ for the complementary group of factors. Moreover, we abusively note the concatenation $(X_I, X_{^cI}) \stackrel{\text{def}}{=} (X_i)_{1 \leq i \leq d}$.

The use of correlation ratio for sensitivity analysis has been proposed by Iman and Hora (1990) and Ishigami and Homma (1990), and independently by Sobol' (1990, 1993). The latter was the most popularized, introducing modifications of correlation ratios of groups of factors to achieve a convenient decomposition of the total variance of the phenomenon, provided that the factors are independent; these are the Sobol' indices. While they are theoretically interesting for studying specific interactions of factors, in practice the most useful sensitivity indices are the *first-order indices* and the *total-order indices*. The former tend to evaluate the influence of a group of factor $I$ on its own and is simply $\eta^2(X_I, Y)$, and the latter incorporate all possible interactions with other factors, defined as $1 - \eta^2(X_{^cI}, Y)$.

Estimation of correlation ratio can be expensive because it involves the term $E(E[Y \mid X_I]^2)$. Most common efficient estimators develop the square conditional expectation as the product $E[f(X_I, X_{^cI}) \mid X_I] E[f(X_I, X'_{^cI}) \mid X_I]$ where $(X_I, X'_{^cI})$ is distributed identically to $(X_I, X_{^cI})$, which in turn is $E[f(X_I, X_{^cI}) f(X_I, X'_{^cI}) \mid X_I]$, provided that $X_{^cI}$ and $X'_{^cI}$ are independent conditionally to $X_I$. In practice, this is ensured when the *input factors are independent*. The expectation of the last expression is nothing but $E(f(X_I, X_{^cI}) f(X_I, X'_{^cI}))$, which is now easier to handle. Typical estimator consists in drawing $2n$ independent observations $(X_I^{(j)}, X_{^cI}^{(j)})_{1 \leq j \leq 2n}$ distributed as $(X_I, X_{^cI})$, and evaluating the model at specifically chosen factors combinations, typically

$$E(E[Y \mid X_I]^2)_n \stackrel{\text{def}}{=} \frac{1}{n} \sum_{j=1}^{n} f(X_I^{(j)}, X_{^cI}^{(j)}) f(X_I^{(j)}, X_{^cI}^{(n+j)}). \tag{2.1}$$

This approach, usually referred to as *pick-and-freeze*, has two drawbacks: first, this constrains the *experience design* (the set of points at which the model must be observed or computed), and second, the required number of model evaluations grows with the number of factors to be investigated.

It is also possible to estimate directly the conditional expectations, and sum its square over observed values of $X_I$. Conditioning by well chosen ranges of the input factors might improve the estimation, but more involved techniques should be considered such as local polynomial regression as described by Da Veiga et al. (2009). Getting reasonable estimations with this approach requires more observations than the previous one, but it is rid of the aforementioned drawbacks, and do not assume independence of the factors.

Another estimation approach worth mentioning relies on sampling the input factors at different, carefully chosen frequencies, and isolating their influence by Fourier decomposition of the resulting phenomenon sample. This is called *Fourier amplitude sensitivity test*. Mara (2009) sums up different improvements proposed over time, allowing to reduce the number of observations required for computing several correlation ratios. However, the experience design is strongly constrained, and correlation ratios of different group of factors require in general different samples. Typically, it is possible to compute all first-order indices from only one sample, but a different sample must be observed for each total-order index.

Let us mention here that we propose yet another estimation approach inspired by *randomized maximum correlation* developed along § 3.5.2. It is similar in spirit to the approach of Da Veiga et al. (2009), in the sense that it approximates the conditional expectation with help of prescribed nonlinear functionals, but it uses different tools. We emphasize the fact that it does not constrain the experience design, that the correlation ratio of all possible groups of factors can be computed from the same random sample, and that it does not rely on statistical independence of the input factors.

Note finally that the correlation ratio is properly defined only for scalar variables. However, a straightforward multidimensional extension is obtained for $\mathcal{Y} \subseteq \mathbb{R}^p$ by summing the correlation ratios of each coordinate weighted by its contribution in the total variance, that is to say

$$\eta^2(X, Y) \stackrel{\text{def}}{=} \frac{\sum_{j=1}^{p} V(Y_j) \eta^2(X, Y_j)}{\sum_{j=1}^{p} V(Y_j)} = \frac{\sum_{j=1}^{p} V(E[Y_j \mid X])}{\sum_{j=1}^{p} V(Y_j)},$$

as studied by Gamboa et al. (2014).

## 3 Sensitivity Analysis with Dependence Measures

Sensitivity analysis based on correlation ratio as described above is fairly general and can be readily adapted for target and conditional analysis, as we propose later in §§ 4.2.1 and 4.2.2. However, several weaknesses can be pointed out.

First, accurate estimation is known for requiring many observations. In addition, although statistical independence implies zero correlation ratio, some variables can be significantly related and yet their correlation ratio be zero as well (as shown by the functional of Ishigami and Homma, 1990, see our numerical illustration § 5.1.2); in such case, one must resort to total-order indices to identify a relation. More generally, the statistical variance of the phenomenon might not be the most representative mode of variation. Finally, the above extension for multidimensional phenomenon might not be satisfying. In the classical probabilistic framework, we believe with Da Veiga (2015) that a more general and more versatile notion of sensitivity of a phenomenon to a group of factors can be captured by the notion of statistical dependence.

First, we review available nonparametric dependence measures. This provides a clearer view of these tools, allowing to comment on various approaches which have been followed in the specific context of sensitivity analysis. Finally, we detail those that we intent to study, making explicit both theoretical definitions and the estimators that we use.

## 3.1 Review on Nonparametric Dependence Measures

We review here the key concepts and tools developed for nonparametric dependence measures.

### 3.1.1 Brief History and Classification

The majority of nonparametric dependence measures between two random variables $X$ and $Y$ rely on the measure of the dissimilarity between their joint probability distribution $P_{X,Y}$ and the product of their marginals $P_X \otimes P_Y$; they are equal if, and only if, $X$ and $Y$ are independent. This strategy dates back at least to Hoeffding (1948), who compares the cumulative distribution functions $F_{X,Y}$ and the separable product $F_X F_Y$ in the spirit of the tests of Kolmogorov–Smirnov (with the supremum norm) and Cramér–Von Mises (with square $L_2$-norm). This approach is further analyzed by Blum et al. (1961); multidimensional extensions are more recently considered by Fernández and González-Barrios (2004).

Instead of using cumulative distribution functions for comparing probability distributions, Rosenblatt (1975) compares *probability density functions* through weighted square $L_2$-norm. Multidimensional extension and further analysis is done, for instance by Anderson et al. (1994) and Ahmad and Li (1997). In turn, weighted square $L_2$ distance of *characteristic functions* is considered by Feuerverger (1993) who mentions similarity with the work of Rosenblatt (1975); later Székely et al. (2007) writes a multidimensional extension under the name *distance covariance*. Other notable approaches based on characteristic functions are proposed by Kankainen and Ushakov (1998) and later by Achard et al. (2003). Interestingly, all these *quadratic dependence measures* share the same estimators, involving *kernel functions* over the ranges of $X$ and $Y$, being either means of density estimation, weight functions or test functions, depending on the approach considered. Gretton et al. (2005) and Diks and Panchenko (2007) independently generalize and give more explicit role to such kernels. In particular, the former rederive the measure in the context of *reproducing kernel Hilbert spaces* (see § 3.3 below), as the Hilbert–Schmidt norm of a generalized cross-covariance operator of $X$ and $Y$, and call it *Hilbert–Schmidt independence criterion*.

Another class of dissimilarity measures between probability distributions is offered by *Csiszár (1972) divergences*. Stemming from *mutual information* as defined by Shannon (1948), extended as *information divergence* between distributions by Kullback and Leibler (1951), the general form (see § 3.4) is also considered independently by Morimoto (1963) and Ali and Silvey (1966). Encompassing several older dependence measures such as Pearson or Neyman $\chi^2$, its use as a dependence measure is in particular advocated by Joe (1989), who underlines the fact that it is suitable for random variables which are multidimensional, but also both continuous and categorical.

Another line of dependence measures does not rely on dissimilarity between joint and product distributions but rather on the concept of correlation. Certainly among the oldest statistics are the correlation coefficients of Pearson (linear), Spearman (on the ranks) and Kendall (on the sign of differences); these can only capture restricted forms of dependence, namely linear for the first one and monotonous for the two others. The *maximum correlation coefficient*, defined as the supremum of the linear correlation coefficient between (measurable, of positive finite variance) transformations of $X$ and $Y$, alleviates this drawback. Introduced originally by Gebelein (1941), its desirable theoretical properties are discussed by Rényi (1959). Practical estimation might be cumbersome even in the simple bivariate case (see the procedure of Breiman and Friedman, 1985). However, by restricting the space of transformations to a finite dimensional setting, it reduces to a simple *canonical correlation*, developed earlier and in a different context by Hotelling (1936). This is exploited by Lopez-Paz et al. (2013) who consider random projections (see § 3.5 below), yielding a form of *randomized maximum correlation*. Let us also mention another generalization of correlation by translation-invariant kernels proposed by Rao et al. (2011), which can also be interpreted as a restricted form of maximum correlation coefficient.

Below, §§ 3.3–3.5, we detail specifically kernel quadratic dependence measures, Csiszár divergence dependence measures and randomized maximum correlation. In addition, we give in § 3.6 a few words on more general classes of measures of dissimilarity between distributions, namely *integral probability metric* and *optimal transport cost*, which could suggest new dependence measures also well adapted to our purpose. Before that, we would like to give a few words about an increasingly popular and useful concept in the study of statistical dependence, the *copula*, which can complement and enhance each considered dependence measure.

### 3.1.2 On the Use of Copulas

Statistical dependence between random variables is a characteristic of their joint distribution; transforming each marginal bijectively would essentially yields the same joint distribution, only on a different space of values. For real random vectors, copulas capture well this idea. The joint distribution of $(X_i)_{1 \leq i \leq d}$ is fully characterized by its cumulative distribution function $F_{(X_i)_{1 \leq i \leq d}} \colon \mathbb{R}^d \to [0,1] \colon x \mapsto \mathrm{P}\bigl(\cap_{i=1}^d \{X_i \leq x_i\}\bigr)$. If each $X_i$ is continuous, then each random variables $F_{X_i}(X_i)$ is uniformly distributed over $[0,1]$ and the copula of $(X_i)_{1 \leq i \leq d}$ is nothing but the cumulative distribution function of $\bigl(F_{X_i}(X_i)\bigr)_{1 \leq i \leq d}$. More generally, it is a cumulative distribution function $C \colon [0,1]^d \to [0,1]$ with uniform marginals, such that for all $x \in \mathbb{R}^d$, $F_{(X_i)_{1 \leq i \leq d}}(x) = C\bigl(F_{X_1}(x_1), \cdots, F_{X_d}(x_d)\bigr)$.

For precise definition and many applications, we refer the reader to the introduction and collected results by Jaworski et al. (2010). We only clarify below some simple aspects, which present the most interest in our context. First, it fully captures the statistical dependences, and most dependence measures between random variables can be easily expressed as a functional of their copula. As pointed by Schmid et al. (2010), this turns out to be a convenient way of defining multivariate versions of classical bivariate dependence measures, such as Spearman and Kendall correlation coefficient mentioned above.

Second, observe that for a real random variable $X$, $F_X(X)$ is invariant under strictly increasing transformations of $X$; in particular, translations or positive scalings. Even for dependence measures which are readily invariant under such transformations on the marginals, their estimators are usually not. Thus, it might be preferable to apply this *copula transform* of the marginals prior to estimation. A desirable consequence is that this provides a normalization of the variable values. This helps addressing our concern on comparisons of dependence measures applied to factors of different natures, see § 3.2.1. Moreover, knowing that the data are falling into the unit cube allows to tailor some estimators accordingly. Both advantages are discussed in more details for the different dependence measures we consider below.

In any case it is worth mentioning here that even though $F_X(X)$ is not directly observable,

observations thereof can be well approximated by composing observations of $X$ with the *empirical distribution functions* deduced from them. Given $(X^{(i)})_{1 \leq i \leq n}$ independent observations distributed identically to $X$, it is $F_{X,n} \stackrel{\text{def}}{=} \frac{1}{n} \sum_{i=1}^{n} 1_{]-\infty, X^{(i)}]}$. Most, if not all, applications of copula transforms, use this approximation. Thus in the following, copula transformed versions of the estimators simply mean replacing the original observations by the empirical copula transformation $(U^{(i)})_{1 \leq i \leq n} \stackrel{\text{def}}{=} (F_{X,n}(X^{(i)}))_{1 \leq i \leq n}$ over all the marginals of the involved random variables.

## 3.2 Dependence Measures for Sensitivity Analysis

Let us give some general considerations before referencing notable uses of dependence measures for sensitivity analysis in the literature.

### 3.2.1 General Considerations

Rényi (1959) states that, among others, a desirable property of a dependence measure DM between two variables $X$ and $Y$ is that it reaches its minimum possible value if and only if $X$ and $Y$ are statistically independent. If used in the right statistical framework, this can be used for *screening*, discarding the factors $X$ which are truly noninfluential on the phenomenon $Y$, detected by low values of $\text{DM}(X, Y)$.

Then, in the hope of using dependence measures for *ranking*, it is necessary that the most important factors exhibit the highest statistical dependence with the phenomenon. This is akin to a *monotonicity* property, although the very concept of being "more important" remains here to be defined. For instance, a factor $X'$ containing more information than a factor $X$ should present more dependence with the phenomenon $Y$. In measure-theoretic formulation, this is expressed by $\mathfrak{F}(X) \subset \mathfrak{F}(X') \implies \text{DM}(X, Y) \leq \text{DM}(X', Y)$, where $\mathfrak{F}(X)$ denotes the $\sigma$-field induced by the random variable $X$. At the very least, one wants $\text{DM}(X_1, Y) \leq \text{DM}((X_1, X_2), Y)$ when pooling factors together.

Such monotonicity property can be easily established for the correlation ratio and the maximum correlation coefficient, and with some care for Csiszar divergence dependence measures; kernel quadratic dependence measures are more delicate in that respect. But this is only for the theoretical quantities; actually, the difficulty arises as soon as the mathematical objects, involved either in the definition or in the estimation of the dependence measure, depend on the spaces in which the studied factors lie. It can be noted that, of all the sensitivity measures we deal with in this work, only the pick-and-freeze estimator, equation 2.1, does not suffer from this effect.

Special care must thus be taken when comparing the importance of factors of different natures or of different dimensions with dependence measures. Fortunately, applying some normalization, for instance according to the highest possible value of the dependence measure considered if available, or resorting to *copula transforms* (see § 3.1.2), can enhance comparisons between factors of different nature and scalings. Nevertheless, this is not systematic and usually cannot remedy poor estimations in high dimension. For more considerations on the sensitivity of dependence measures to the dimension, we refer again the reader to Da Veiga (2015), who notably mentions the possibility of designing multidimensional factors selection procedures, involving only dependence measures applied to unidimensional marginals, inspired by feature selection methods for machine learning.

### 3.2.2 Literature Review

Many different tools have been proposed for the purpose of sensitivity analysis, and some can be recast in our rough classification of dependence measures, § 3.1.1; although usually with some substantial modifications.

First, let us set aside all the tools relying on correlation coefficient, linear regressions, correlation ratios as described in § 2, or even simple transformations thereof as discussed for instance in Saltelli and Sobol' (1995). As already pointed out, those can detect only restricted forms of dependence, in contrast to tools involving entire distributions. In the sensitivity analysis community, the latter have been often referred to as *moment-independent uncertainty measures*.

One of the earliest work explicitly in that direction is the one of Park and Ahn (1994), who use something close to the mutual information (which is a dependence measure based on Csiszár divergence, see § 3.4.2). More precisely, they use the Kullback–Leibler divergence to compare the distribution of the phenomenon, $P_Y$, to its distribution if the studied factor $X$ undergoes a "distributional change"; think typically to a conditioning, $P_{Y|X \in A}$. Later, Borgonovo (2007) suggests integrating the divergence between original and conditional distribution against the distribution of the factor $X$, yielding exactly the Csiszár divergence dependence measure (this time with the total variation flavor). In both cases, estimating the Csiszár divergence over conditional densities is problematic, and they must call on *ad hoc* parametric density fits. Several authors explore other estimation strategies with various degrees of mathematical rigor (Liu and Homma, 2009; Wei et al., 2013; Luo et al., 2014; Jia, 2014; Da Veiga, 2015; Rahman, 2016). These methods are often designated as *density-based* uncertainty importance measure.

Dependence measures based on cumulative distribution functions have also been considered. Chun et al. (2000) follow this path. The metric they use is peculiar, not comparing directly the cumulative distribution functions but rather the quantile functions (which are their generalized, left-continuous inverse). Moreover, like Park and Ahn before them, they only compare $P_Y$ with a "distributional change" as described above. It is interesting to note that the regional sensitivity analysis of Spear and Hornberger (1980), presented in § 1.2.1 in the context of target sensitivity analysis, actually follows the same logic. In essence, one could use conditioning by each possible value of $Y$, and integrate the resulting Kolmogorov distances against its distribution, to make it a proper dependence measure. Although the role of $X$ and $Y$ are interchanged, this is exactly what is proposed by Baucells and Borgonovo (2013). The estimation procedure that they propose resembles a costly double loop, the inner for estimating the conditioned distributions by a rough kernel method, the outer for integrating. Recently, Gamboa et al. (2017) study the analogous dependence measure obtained by replacing the Kolmogorov distance by a Cramér–Von Mises distance weighted by the distribution of $Y$. Observing that

$$\mathrm{E}\left[\int_{\mathcal{Y}} \left(F_Y(y) - F_{Y|X}(y)\right)^2 \, \mathrm{d}P_Y(y)\right] = \int_{\mathcal{Y}} \mathrm{E}\left[\left(\mathrm{E}[1_{\{Y \leq y\}}] - \mathrm{E}[1_{\{Y \leq y\}} \mid X]\right)^2\right] \mathrm{d}P_Y(y),$$

and that the expectation within the right-hand side integral is nothing but $\mathrm{V}(\mathrm{E}[1_{Y \leq y} \mid X])$, the authors suggests to estimate it with help of the pick-and-freeze method, described in § 2. Then, the outer integration can be estimated at the cost of only an additional independent sample distributed as $P_Y$. Indeed, all the pick-and-freeze estimates in the integration can be computed from the same set of factors combinations and corresponding phenomenon, only transforming it according to $1_{\{Y \leq y\}}$ for each level $y$. This presents several advantages. First, the authors are able to derive consistency and asymptotic normality of the resulting estimator. Second, in terms of interpretation, this shows how this dependence measure can be seen as a generalization of the correlation ratio. It can be similarly normalized, and when the factors are independent, just like the latter decomposes the variance, the former can be seen as decomposing the dependence among groups of factors. Moreover, as pointed out above, § 3.2.1, the use of pick-and-freeze estimators eases the comparison of factors of different natures and dimensions. Unfortunately, this comes also with important drawbacks: the estimator rely on factors independence, and the total computational cost grows with the number of factors to study.

In spite of strong similarities, none of the above dependence measures seem equivalent to the direct application of a distance between the joint $F_{X,Y}$ and product $F_X F_Y$ cumulative distribution functions. As these are well approximated by their empirical counterparts $F_{X,Y,n}$ and $F_{X,n} F_{Y,n}$, this suggests yet another dependence measure with a convenient estimator. We do not explore it further in the present study for the sake of brevity, but it certainly have its own interest and should be considered in future works.

As for the whole class of kernel quadratic dependence measures, their only mention in the context of sensitivity analysis that we are aware of is given by Da Veiga (2015), complemented by numerical illustrations of De Lozzo and Marrel (2016). As far as we know, maximum correlation and the likes never appeared in this context, except for one mention in the latter article. In the following, we develop these two, together with the Csiszár divergence dependence measure which we extend theoretically. Note that our choice is mainly guided by ease of implementation (notably the possibility of writing estimators as empirical expectations), aim for generality (factors and phenomenon of any nature and dimension), good invariance properties (as discussed in various references given in § 3.1.1), and ease of adaptation for target and conditional sensitivity analysis.

## 3.3 Kernel Quadratic Dependence Measure

We give here the interpretation of the kernel quadratic dependence measure through kernel embeddings of probability distributions in reproducing kernel Hilbert spaces (see for instance Berlinet and Thomas-Agnan, 2003, Chapter 4). We then derive the dependence measure in our setting, specify a typical estimator, and give some considerations for sensitivity analysis.

### 3.3.1 Kernel Distance

Let $\mathcal{Z}$ be a generic topological space, and $\mathcal{H}_k \subset \mathbb{R}^{\mathcal{Z}}$ a (separable) reproducing kernel Hilbert space induced by a *positive definite kernel* $k \colon \mathcal{Z}^2 \to \mathbb{R}$. If P is a probability distribution over $\mathcal{Z}$, and $k$ is measurable such that $\int k(z,z') \, \mathrm{d}P(z) \, \mathrm{d}P(z') < +\infty$, the linear form $\mathcal{H}_k \to \mathbb{R}$ defined through P as $\phi \mapsto \int \phi \, \mathrm{d}P$ is bounded, and has thus a representer $\mu_k(P) \in \mathcal{H}_k$. Such representer is the *kernel embedding* of the distribution P in $\mathcal{H}_k$, and can be expressed as $z \mapsto \int k(z,z') \, \mathrm{d}P(z')$.

Now, inner products of kernel embeddings can be expressed as expectations of the kernel $k$; more precisely, if Q is another probability distribution over $\mathcal{Z}$ with the above properties, then $\langle \mu_k(P), \mu_k(Q) \rangle_{\mathcal{H}_k} = \int k(z,z') \, \mathrm{d}P(z) \, \mathrm{d}Q(z')$. The associated norm $\|\mu_k(P) - \mu_k(Q)\|_{\mathcal{H}_k}$ is the *kernel distance*, also referred to as *maximum mean discrepancy*. It is especially relevant when the map $P \mapsto \mu_k(P)$ is injective, in which case the kernel distance is zero if, and only if, the distributions are the same; a kernel enjoying this property is called *characteristic*. Typical characteristic kernels are the *Gaussian kernel*, when $\mathcal{Z}$ is a normed vector space, $(z,z') \mapsto \exp(-\|z-z'\|^2/2\sigma^2)$ for some parameter $\sigma^2 \in \mathbb{R}$ dependent in practice on the data; and the *categorical kernel*, when $\mathcal{Z}$ is a finite set, $(z,z') \mapsto 1$ if $z = z'$, 0 otherwise.

### 3.3.2 Quadratic Dependence Measure

Back to our setting, a measure of the dependence between $X$ and $Y$ is thus defined by the kernel distance between $P_{X,Y}$ and $P_X \otimes P_Y$. These are probability distributions over the space $\mathcal{X} \times \mathcal{Y}$; a useful particular case arises when the kernel $k$ is separable as $((x,y),(x',y')) \mapsto k_{\mathcal{X}}(x,x')k_{\mathcal{Y}}(y,y')$, where $k_{\mathcal{X}}$ and $k_{\mathcal{Y}}$ are positive definite kernels over $\mathcal{X}$ and $\mathcal{Y}$, respectively. In that case, the square kernel distance $\|\mu_{k_{\mathcal{X}}k_{\mathcal{Y}}}(P_{X,Y}) - \mu_{k_{\mathcal{X}}k_{\mathcal{Y}}}(P_X \otimes P_Y)\|^2_{\mathcal{H}_{k_{\mathcal{X}}k_{\mathcal{Y}}}}$ reads as

$$E\big(k_{\mathcal{X}}(X,X')k_{\mathcal{Y}}(Y,Y')\big) + E\big(k_{\mathcal{X}}(X,X')\big)E\big(k_{\mathcal{Y}}(Y,Y')\big) - 2E\big(k_{\mathcal{X}}(X,X')k_{\mathcal{Y}}(Y,Y'')\big), \quad (3.1)$$

provided that $(X', Y')$ is independent of, and distributed identically to, $(X, Y)$, and $Y''$ is independent of $X, Y, X', Y'$ and distributed identically to $Y$. This yields the kernel quadratic dependence measure, which we note $\mathrm{QDM}_{k_{\mathcal{X}}, k_{\mathcal{Y}}}(X, Y)$; it is also called Hilbert–Schmidt independence criterion by Gretton et al. (2005).

A straightforward estimator, given $(X^{(i)}, Y^{(i)})_{1 \le i \le n}$ independent observations distributed identically to $(X, Y)$, is[4]

$$\begin{aligned}
\mathrm{QDM}_{k_{\mathcal{X}}, k_{\mathcal{Y}}}(X, Y)_n &\stackrel{\text{def}}{=} \frac{1}{n^2} \sum_{i,j=1}^n k_{\mathcal{X}}\big(X^{(i)}, X^{(j)}\big) k_{\mathcal{Y}}\big(Y^{(i)}, Y^{(j)}\big) \\
&\quad + \frac{1}{n^2}\bigg(\sum_{i,j=1}^n k_{\mathcal{X}}\big(X^{(i)}, X^{(j)}\big)\bigg) \frac{1}{n^2}\bigg(\sum_{i,j=1}^n k_{\mathcal{Y}}\big(Y^{(i)}, Y^{(j)}\big)\bigg) \\
&\quad - \frac{2}{n} \sum_{i=1}^n \bigg(\frac{1}{n} \sum_{j=1}^n k_{\mathcal{X}}\big(X^{(i)}, X^{(j)}\big)\bigg)\bigg(\frac{1}{n} \sum_{j=1}^n k_{\mathcal{Y}}\big(Y^{(i)}, Y^{(j)}\big)\bigg),
\end{aligned}$$

which should be put under the following handier form for practical implementation,

$$\frac{1}{n^2} \sum_{i,j=1}^n \bigg(k_{\mathcal{X}}\big(X^{(i)}, X^{(j)}\big) - \frac{1}{n} \sum_{\ell=1}^n k_{\mathcal{X}}\big(X^{(i)}, X^{(\ell)}\big)\bigg)\bigg(k_{\mathcal{Y}}\big(Y^{(i)}, Y^{(j)}\big) - \frac{1}{n} \sum_{\ell=1}^n k_{\mathcal{Y}}\big(Y^{(\ell)}, Y^{(j)}\big)\bigg).$$

Note that some authors prefer normalizing with factors $n-1$ and $n-2$, or add some other debiasing modifications, which are of little interest here. The required number of computations grows as $O(n^2)$, which is acceptable in situations where the cost for obtaining each observation is large.

When $\mathcal{X}$ and $\mathcal{Y}$ are normed vector spaces, in addition to typical Gaussian and categorical kernels given above, we would like to mention the particular case of the distance covariance of Székely et al. (2007). It was shown later by Sejdinovic et al. (2013) that it is indeed a form of kernel quadratic dependence measure, with the appropriate kernels $k_{\mathcal{X}} : (x, x') \mapsto \frac{1}{2}(\|x\| + \|x'\|) - \|x - x'\|$, and similarly for $Y$.

Székely et al. (2007) propose to normalize their dependence measure in the spirit of the linear correlation coefficient, yielding the *distance correlation*, which is scale-invariant. When Da Veiga (2015) highlights the potential use of the quadratic dependence measure for sensitivity analysis, he also advocates for such normalization, leading to the sensitivity index

$$\overline{\mathrm{QDM}}_{k_{\mathcal{X}}, k_{\mathcal{Y}}}(X, Y) \stackrel{\text{def}}{=} \frac{\mathrm{QDM}_{k_{\mathcal{X}}, k_{\mathcal{Y}}}(X, Y)}{\sqrt{\mathrm{QDM}_{k_{\mathcal{X}}, k_{\mathcal{X}}}(X, X)} \sqrt{\mathrm{QDM}_{k_{\mathcal{Y}}, k_{\mathcal{Y}}}(Y, Y)}},$$

and similarly for its *plug-in* estimator $\overline{\mathrm{QDM}}_{k_{\mathcal{X}}, k_{\mathcal{Y}}}(X, Y)_n$. This should allow for direct comparisons of dependence measures between a phenomenon and group of factors of different nature.

---

[4] We omit the fact that the parameters of the kernel might depend on the data.

Alternatively, Póczos et al. (2012) recommend applying kernel quadratic dependence measures on the copula transforms. For multidimensional factors, it might be useful to combine both normalizations.

### 3.4 Csiszár Divergence Dependence Measure

We now turn to Csiszár divergences. As previously, we explain the general discrepancy measures between distributions before detailing its use for dependence measures and sensitivity analysis.

#### 3.4.1 Csiszár Divergence

Considering again P, Q two probability distributions over a generic space $\mathcal{Z}$, a *Csiszár divergence* between P and Q can be defined as

$$\mathrm{div}_\phi(\mathrm{P},\mathrm{Q}) \stackrel{\mathrm{def}}{=} \int \phi\left(\frac{\mathrm{dP}}{\mathrm{dQ}}\right) \mathrm{dQ},$$

where $\frac{\mathrm{dP}}{\mathrm{dQ}}$ is the Radon–Nikodym derivative of P with respect to Q, and $\phi\colon \mathbb{R}_+ \to \mathbb{R} \cup \{+\infty\}$ is a convex function vanishing at unity. Using Jensen inequality, $\mathrm{div}_\phi(\mathrm{P},\mathrm{Q}) \geq \phi\left(\int \mathrm{dP}\right) = \phi(1) = 0$, hence takes values in $\mathbb{R}_+ \cup \{+\infty\}$. This can be conveniently extended to cases where P is not dominated by Q. Lebesgue decomposition theorem ensures $\mathrm{P} = \mathrm{P}_\ll + \mathrm{P}_\perp$ where $\mathrm{P}_\ll$ and $\mathrm{P}_\perp$ are respectively absolutely continuous and singular with respect to Q, and we can define

$$\mathrm{div}_\phi(\mathrm{P},\mathrm{Q}) \stackrel{\mathrm{def}}{=} \int \phi\left(\frac{\mathrm{dP}_\ll}{\mathrm{dQ}}\right) \mathrm{dQ} + \phi^*(0) \mathrm{P}_\perp(\mathcal{Z}), \qquad (3.2)$$

where $\phi^*\colon \mathbb{R}_+ \to \mathbb{R} \cup \{+\infty\}\colon t \mapsto t\phi(1/t)$ if $t > 0$ and $\phi^*(0) = \lim_{t \to 0} \phi^*(t)$, with the usual integration convention $0 \times \infty = 0$. Note that the role of this second term is to handle properly the domain where, in essence, Q is zero but P is not, that is where the derivative tends to infinity but is integrated over a set of measure zero. It can be noted also that $\mathrm{div}_\phi(\mathrm{P},\mathrm{Q}) = \mathrm{div}_{\phi^*}(\mathrm{Q},\mathrm{P})$; we refer to Liese and Vajda (2006) for a complete treatment.

Notable examples include:
- the *Kullback–Leibler divergence* with $\phi\colon t \mapsto t\log(t)$, $\phi^*\colon t \mapsto -\log(t)$ and $\phi^*(0) = +\infty$;
- the *total variation distance* with $\phi\colon t \mapsto |t-1|$, $\phi^* = \phi$ and $\phi^*(0) = 1$;
- the square *Hellinger distance* with $\phi\colon t \mapsto (\sqrt{t}-1)^2$, satisfying also $\phi^* = \phi$ and $\phi^*(0) = 1$;
- the *Pearson $\chi^2$* with $\phi\colon t \mapsto (t-1)^2$, $\phi^*\colon t \mapsto \frac{(t-1)^2}{t}$ and $\phi^*(0) = +\infty$;

and all their dual, by switching the role of $\phi$ and $\phi^*$: respectively *reverse* Kullback–Leibler (note that $\lim_{t \to 0} t\log(t) = 0$), total variation and square Hellinger distances again, and *Neyman $\chi^2$* ($\lim_{t \to 0}(t-1)^2 = 1$).

#### 3.4.2 Application to Dependence Measure and Estimation for Sensitivity Analysis

Again, one can measure dependence between $X$ and $Y$ by measuring discrepancy between $\mathrm{P}_{X,Y}$ and $\mathrm{P}_X \otimes \mathrm{P}_Y$ with help of Csiszár divergences. The famous special case of the mutual information is obtained with the Kullback–Leibler divergence, that is $\mathrm{div}_{t \mapsto t\log(t)}\left(\mathrm{P}_{X,Y}, \mathrm{P}_X \otimes \mathrm{P}_Y\right)$ or equivalently $\mathrm{div}_{-\log}\left(\mathrm{P}_X \otimes \mathrm{P}_Y, \mathrm{P}_{X,Y}\right)$. We refer to our literature review, § 3.2.2, for other specific uses and estimations for sensitivity analysis.

Estimations of Csiszár divergences have been studied in many contexts, often focused on specific versions defined by a given function $\phi$ or on specific knowledge about the involved distributions. In order to devise a general tool, we choose in the current work to rely on nonparametric estimations of the Radon–Nikodym derivatives. *In practice, we consider factors which are either continuous* (with respect to Lebesgues measure) *or discrete* (categorical). Densities at specific points can be estimated through *kernel* or *nearest-neighbors* methods, see for instance the monograph of Silverman (1986). Probabilities are estimated by empirical frequencies. Both can be combined if necessary.

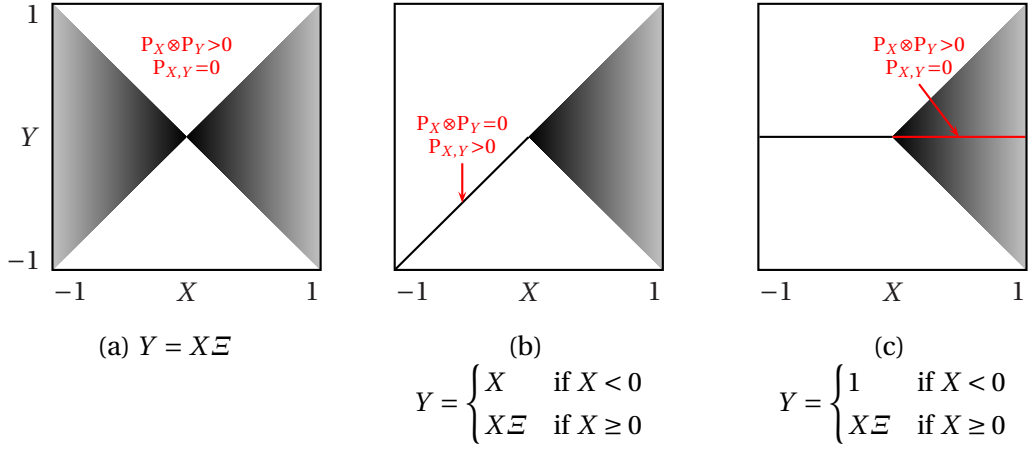

Figure 1: Illustration of cases of mutual singularities of $P_X \otimes P_Y$ and $P_{X,Y}$ in some functional relationships $Y = f(X, \Xi)$. In each case, $X$ and $\Xi$ are independently and uniformly distributed over $[-1,1]$. The joint distribution is represented in scales of gray, the darker the higher probability. With respect to the two-dimensional Lebesgues measure, smooth areas represent absolutely continuous parts, while black lines schematize singular parts.

Let us give special considerations concerning the singular parts appearing in equation 3.2. Under our assumptions, we identify three cases of mutual singularities between $P_X \otimes P_Y$ and $P_{X,Y}$, illustrated in figure 1. The first one appears in most functional relationships, where some values of $Y$ never occurs with some values of $X$; this is illustrated on figure 1(a), and concerns also each white area in figures 1(b)–(c). The second and third arise when $Y$ takes on continuous values, but get a discrete distribution conditionally to some values of $X$. In figures 1(b)–(c), some values of $X$ completely determine $Y$, leading to different singularities depending on if $Y$ varies continuously with $X$ or not.

In any case, we do not try to estimate the domains of continuity and singularities, but rely instead on the regularizing effect of kernel density estimates. In the same time, integration against probability measures is performed through Monte Carlo approach. Altogether, we can hope that the resulting estimates get close to the actual Csiszár divergence, equation 3.2. Noting that the inclusion $\mathrm{ran}(X,Y) \subset \mathcal{X} \times \mathcal{Y}$ is usually strict as is illustrated on figure 1, we distinguish two cases depending on the integration domain.

First, we consider the full *Csiszár divergence dependence measure* simply as $\mathrm{CDM}_\phi(X,Y) \stackrel{\mathrm{def}}{=} \mathrm{div}_\phi(P_{X,Y}, P_X \otimes P_Y)$. It is expected to be difficult to estimate, because the integration domain $\mathcal{X} \times \mathcal{Y}$ is the largest, and is *not* an expectation of a function of the *joint* variable $(X,Y)$. A natural estimator based on kernel density estimate is

$$\mathrm{CDM}_\phi(X,Y)_{k_\mathcal{X}, k_\mathcal{Y}, n} \stackrel{\mathrm{def}}{=} \frac{1}{n^2} \sum_{i,j=1}^{n} \phi\left( \frac{\frac{1}{n}\sum_{\ell=1}^n k_{\mathcal{X},\mathcal{Y}}\left((X^{(i)}, Y^{(j)}), (X^{(\ell)}, Y^{(\ell)})\right)}{\left(\frac{1}{n}\sum_{\ell=1}^n k_\mathcal{X}(X^{(i)}, X^{(\ell)})\right)\left(\frac{1}{n}\sum_{\ell=1}^n k_\mathcal{Y}(Y^{(j)}, Y^{(\ell)})\right)} \right), \quad (3.3)$$

where $k_\mathcal{X}$, $k_\mathcal{Y}$ and $k_{\mathcal{X},\mathcal{Y}}$ are the kernels used for estimating densities or probabilities; typically (normalized) Gaussian and categorical, respectively. Defining $k_{\mathcal{X},\mathcal{Y}}$ as the separable product $\big((x, y), (x', y')\big) \mapsto k_{\mathcal{X}}(x, x') k_{\mathcal{Y}}(y, y')$ might ease the computation of the numerator, but it still has a total cost growing as $O(n^3)$, which is very expensive. Nearest-neighbors density are in general even more expensive.

In order to alleviate the computational burden, we could consider summing over only $\lfloor n/2 \rfloor$ truly independent pairs $\big(X^{(2i-1)}, Y^{(2i)}\big)_{1 \leq i \leq \lfloor n/2 \rfloor}$, in place of all the $n^2$ possible pairs $\big(X^{(i)}, Y^{(j)}\big)_{1 \leq i,j \leq n}$. We believe however that the space $\mathcal{X} \times \mathcal{Y}$ would then be too scarcely sampled for accurate estimation. Instead, we prefer to restrict the integration over the more meaningful domain $\operatorname{ran}(X, Y)$. We thus propose to modify the Csiszár divergence as the *support Csiszár divergence* defined as

$$\operatorname{sdiv}_\phi(P, Q) \stackrel{\text{def}}{=} \int \phi\left(\frac{dP_\ll}{dQ}\right) dQ + \phi^*(0) P_\perp\big(\operatorname{supp}(Q)\big), \tag{3.4}$$

where $\operatorname{supp}(Q)$ is the *support of the measure* Q. In other words, we plainly and simply drop singularity of P with respect to Q outside its support. Good general properties of the Csiszár divergence, in particular nonnegativity and identity of indiscernibles, is ensured for $\phi$ positive over $[0, 1]$ (see proposition A.3); this is a mild limitation in regard to the functionals $\phi$ usually considered, as described above.

This allows us to define the *support* Csiszár divergence dependence measure $\operatorname{sCDM}_\phi(X, Y)$ as $\operatorname{sdiv}_\phi\big(P_X \otimes P_Y, P_{X,Y}\big)$. It can be noted that the kind of singularity which is ignored is illustrated on figure 1(a), while the two others, (b) and (c), can be approximated as expectations of functions of the joint variables $(X, Y)$. The corresponding estimator based on kernel density estimate is

$$\operatorname{sCDM}_\phi(X, Y)_{k_\mathcal{X}, k_\mathcal{Y}, n} \stackrel{\text{def}}{=} \frac{1}{n} \sum_{i=1}^{n} \phi\left(\frac{\big(\frac{1}{n}\sum_{j=1}^{n} k_\mathcal{X}(X^{(i)}, X^{(j)})\big)\big(\frac{1}{n}\sum_{j=1}^{n} k_\mathcal{Y}(Y^{(i)}, Y^{(j)})\big)}{\frac{1}{n}\sum_{j=1}^{n} k_{\mathcal{X}, \mathcal{Y}}\big((X^{(i)}, Y^{(i)}), (X^{(j)}, Y^{(j)})\big)}\right), \tag{3.5}$$

which has now a computational cost of $O(n^2)$, just as for the kernel quadratic dependence measure. Such an estimator has been mentioned by several authors (see § 3.2.2), without further justification.

Although variation range of the Csiszár divergence dependence measures can be controlled, sensitivity measures based on them are in general not directly comparable from one factor to another. A Csiszár divergence with a given function $\phi$ might behave differently on continuous and discrete probability measures. Moreover, quality of kernel density estimates are known to be sensitive to the ambient dimension. With only a few points in high dimension, redundancy of information is difficult to identify, usually leading to overestimation of the dependence. Unfortunately, normalization is not as natural as for the kernel quadratic dependence measure which derives from a square norm. Consider moreover that, for instance, $\operatorname{CDM}_\phi(X, X)$ and $\operatorname{sCDM}_\phi(X, X)$ might be infinite.

In fact, normalization of the mutual information is discussed in details by Joe (1989), where different, noncompatible normalization schemes are proposed, depending on whether the variables are continuous or categorical. To our knowledge, such discussion has not even been raised for general Csiszár divergences. We propose to normalize *the estimators* as

$$\overline{\operatorname{CDM}}_\phi(X, Y)_{k_\mathcal{X}, k_\mathcal{Y}, n} \stackrel{\text{def}}{=} \frac{\operatorname{CDM}_\phi(X, Y)_{k_\mathcal{X}, k_\mathcal{Y}, n}}{\operatorname{CDM}_\phi(X, X)_{k_\mathcal{X}, k_\mathcal{X}, n}},$$

and similarly for $\overline{\operatorname{sCDM}}(X, Y)_{k_\mathcal{X}, k_\mathcal{Y}, n}$, having been careful of using $\phi$ positive over $[0, 1]$ to ensure positivity of the estimated quantities. It can be seen as a rough generalization of the normalization proposed by Joe (1989) for mutual information of categorical variables, because sCDM$_{-\log}(X,X)$ is in that case the *Shannon entropy* of $X$.

In addition, the use of copula transforms can be helpful. In theory, Csiszár divergences are already invariant under bijective transformations, but nevertheless their estimators might be sensitive to scaling. Blumentritt and Schmid (2012) show the advantage of using copula transformed for estimation of mutual information, by taking care of adapting the estimation tools (kernel, nearest-neighbors radius, etc.) to the unit cube. Let us mention at this occasion that the same authors propose a normalization of the mutual information according to the dimension. Based on analytical observations on Gaussian copulas, it ensures that, if each pair of marginals have the same correlation, increasing the dimension does not change the multivariate dependence measure. This normalization is not relevant in our context, where adding factors correlated with the phenomenon should in fact increase the dependence measure. Also, Nagler and Czado (2016) propose a much more involved application of copula theory to density estimation. This should be considered when the factors dimension becomes prohibitively large, but high computational overhead is expected.

## 3.5 Maximum Correlation

As mentioned in our review of dependence measures § 3.1, the maximum correlation coefficient between $X$ and $Y$ is defined as

$$\sup\left\{\rho\big(\phi_{\mathcal{X}}(X),\phi_{\mathcal{Y}}(Y)\big)\,\bigg|\,\begin{matrix}\phi_{\mathcal{X}}\in\mathbb{R}^{\mathcal{X}}\text{ measurable, }0<\mathrm{V}(\phi_{\mathcal{X}}(X))<+\infty\\ \phi_{\mathcal{Y}}\in\mathbb{R}^{\mathcal{Y}}\text{ measurable, }0<\mathrm{V}(\phi_{\mathcal{Y}}(Y))<+\infty\end{matrix}\right\},$$

where $\rho$ is the linear correlation coefficient $(X,Y)\mapsto\frac{\mathrm{C}(X,Y)}{\sqrt{\mathrm{V}(X)}\sqrt{\mathrm{V}(Y)}}$.

### 3.5.1 Practical Estimation

Estimating such a supremum over all measurable functionals is intractable; one possibility is to restrict the search space to a finite-dimensional space spanned by certain finite families of functionals, $(\phi_{\mathcal{X},j})_{1\leq j\leq k}$, $(\phi_{\mathcal{Y},j})_{1\leq j\leq \ell}$. The supremum becomes then the canonical correlation of Hotelling (1936), $\sup_{\alpha\in\mathbb{R}^k,\beta\in\mathbb{R}^\ell}\rho\Big(\sum_{j=1}^{k}\alpha_j\phi_{\mathcal{X},j}(X),\sum_{j=1}^{\ell}\beta_j\phi_{\mathcal{Y},j}(Y)\Big)$.

This is used by Bach and Jordan (2003) with bases functionals constituted by features of a reproducing kernel Hilbert space mapped from observed data. This is called *kernel canonical correlation*. In that case, the finite-dimensional optimization actually amounts to optimization over the whole, possibly infinite-dimensional Hilbert space. While this can be desirable, the estimation over finite sample data requires some kind of regularization.

In a different approach, Lopez-Paz et al. (2013) propose to choose the bases functionals in a prespecified space of nonlinear functionals, depending continuously on some parameters; the key point being to draw randomly the parameter values. Over an Euclidean space $\mathcal{X}\subseteq\mathbb{R}^p$, the authors typically recommends using nonlinear functionals of the form $\phi_{\theta,b}\colon x\mapsto \sin(\langle x,\theta\rangle+b)$, and drawing a certain number of parameters $\theta\in\mathbb{R}^p$, $b\in\mathbb{R}$ according to normal distributions scaled by the dimension $p$.

Interesting connection with kernel canonical correlation can actually be established for translation-invariant kernels. By Bochner theorem, these are Fourier transforms of certain distributions $(x,x')\mapsto\int e^{i\langle x-x',\theta\rangle}\,\mathrm{d}\mu(\theta)$; in a sense, drawing some parameters $\theta$ according to $\mu$ and summing resulting sinusoidal functionals approximate such kernel. In that respect, note that a normal distribution $\mu$ corresponds to a Gaussian kernel.

Randomly drawing a restricted number of nonlinear functionals has two advantages: first, it prevents overfitting over finite sample data, and second, it eases the computations.

However, the selection of this number, as well as the distribution over the parameters, are only heuristically chosen so far. Finally, in order to ensure shift-invariance and scale-invariance, Lopez-Paz et al. (2013) first apply empirical copula transforms on the data. They call the resulting dependence measure the *randomized dependence coefficient*.

### 3.5.2 Randomized Maximum Correlation

In our work, we consider different sets of data transformations and nonlinear functionals composed in $\phi_{\theta_\mathcal{X}}$ and $\phi_{\theta_\mathcal{Y}}$. Given samples of random observations $(X^{(i)}, Y^{(i)})_{1 \leq i \leq n}$ and random parameters $(\Theta_\mathcal{X}^{(j)})_{1 \leq j \leq k}$, $(\Theta_\mathcal{Y}^{(j)})_{1 \leq j \leq \ell}$,[5]

$$\text{RMC}(X, Y)_{\phi_{\theta_\mathcal{X}}, \phi_{\theta_\mathcal{Y}}, n} \stackrel{\text{def}}{=} \sup_{\alpha \in \mathbb{R}^k, \beta \in \mathbb{R}^\ell} \rho_n \left( \sum_{j=1}^k \alpha_j \left( \phi_{\Theta_\mathcal{X}^{(j)}}(X^{(i)}) \right)_{1 \leq i \leq n}, \sum_{j=1}^\ell \beta_j \left( \phi_{\Theta_\mathcal{Y}^{(j)}}(Y^{(i)}) \right)_{1 \leq i \leq n} \right), \quad (3.6)$$

where $\rho_n$ is the empirical correlation coefficient. This can be computed as a solution of an eigenvalue problem involving the variance-covariance matrices of the $k + \ell$ nonlinear projections (that is to say the transformed observations). If $k$ and $\ell$ are small in comparison to $n$, the cost is dominated by the computations of the last matrices. If no complex transformation is applied, then this is in the order $O((k+\ell)^2 n)$. When applying empirical copula transforms, the observations must be sorted at a cost $O(n \log(n))$, which remains low in comparison to other dependence measures.

Choosing appropriate numbers $k$ and $\ell$ might be delicate. On the one hand, when the set of possible nonlinear functionals is not rich enough, important correlation modes can be missed. On the other hand, when too many functionals are introduced, artificial correlation can be captured due to the finite nature of the observation sample; consider that when $k$ or $\ell$ approach $n$, the nonlinear projections end up spanning almost the whole space $\mathbb{R}^n$, thus the canonical correlation gets close to unity. Altogether, the number of nonlinear projections must increase with the number of observations, while remaining largely dominated by it. Without further investigation, we propose to set it as the square root of the number of observations (leading to an overall computational complexity of $O(n^2)$).

To our knowledge, this effect has not been previously emphasized; for instance Lopez-Paz et al. (2013) set the number of nonlinear projections arbitrarily from trial-and-errors. Nevertheless, this effect can play a significant role when comparing randomized maximum correlations between factors. It even becomes crucial if one is interested in an accurate estimation of the actual maximum correlation coefficient, as is the case when comparing randomized maximum correlation with other sensitivity indices, or using it to estimate correlation ratio (see § 3.5.3 below).

In the latter case, it is also necessary to correct for the well-known bias on the correlation coefficient due to the number of regressors. Most typical in the context of *multiple linear regression*, the formula of Wherry (1931) can be straightforwardly adapted to canonical correlation by replacing the number of regressors by $k + \ell - 1$. This yields the following debiased estimator of the square maximum correlation: $1 - \left(1 - \text{RMC}(X, Y)_n^2\right) \frac{n-1}{n-(k+\ell-1)}$. Many other formulae have been investigated ever since (see for instance the work of Leach, 2006, in the context of canonical correlation), but they do no differ much in the regime $1 \ll k, \ell \ll n$.

For the use in sensitivity analysis, a randomized maximum correlation conveniently lies in [0,1] without need for further normalization. Of course, different functionals are required when comparing factors of different nature; it is even conceivable to design functionals for application to categorical variables. In spite of this, it seems usually possible to get randomized maximum correlations that are comparable to one another, by controlling the overall richness of their respective sets of functionals, through their numbers $k$, their expressions, and the distributions over their parameters $\Theta_{\mathcal{X}}$. In that respect, we underline that the canonical correlation is performed with $k$ vectors, whatever the actual dimension of the considered factor is; this can be used to mitigate the influence of dimensionality.

---

[5]The notation omits the dependence in the numbers $k$ and $\ell$, and the expression omits the fact that the transformations within $\phi_{\Theta_\mathcal{X}}$ and $\phi_{\Theta_\mathcal{Y}}$ might depend on the data; for instance when applying empirical copula transforms.

The influence of the distribution of $\Theta_{\mathcal{X}}$ is less intuitive, but might be worth investigating in the future.

### 3.5.3 Link with Correlation Ratio

Suppose for a while that $\mathcal{Y} \subseteq \mathbb{R}$, with $V(Y) < +\infty$. In such case, we know that the conditional expectation of $Y$ is its *orthogonal projection* over the space of measurable functions of $X$, that is to say $E[Y \mid X] = \phi_{\text{opt}}(X)$ with $\phi_{\text{opt}} \in \arg\min_{\phi:\ V(\phi(X))<+\infty} E\big((Y - \phi(X))^2\big)$. This develops as $E(Y^2) + E(\phi(X)^2) - 2E(Y\phi(X)) = V(Y) + V(\phi(X)) - 2C(Y, \phi(X)) + (E(Y) - E(\phi(X)))^2$. Only the last term depends on $E(\phi(X))$, determining $E(\phi(X)) = E(Y)$. We can thus decompose $\phi_{\text{opt}}$ as $E(Y) + \sigma_{\text{opt}} \phi_1$, where $E(\phi_1(X)) = 0$, $V(\phi_1(X)) = 1$ and $\sigma_{\text{opt}} \in \mathbb{R}_+$. Ignoring the constant $V(Y)$ in the minimization, we get $(\sigma_{\text{opt}}, \phi_1) \in \arg\min_{(\sigma,\phi):\ \sigma \in \mathbb{R}_+, V(\phi(X))=1} \sigma^2 - 2\sigma C(Y, \phi(X))$. Since $\sigma$ is nonnegative, we deduce first that $\phi_1 \in \arg\max_{\phi:\ V(\phi(X))=1} C(Y, \phi(X))$, and finally $\sigma_{\text{opt}} = C(Y, \phi_1(X))$. Altogether, we obtain that $\eta^2(X, Y) = \frac{\sigma_{\text{opt}}^2}{V(Y)} = \frac{C(Y,\phi_1(X))^2}{V(Y)} = \rho^2(Y, \phi_1(X))$. By the definition of $\phi_1$ and since the linear correlation coefficient is scale-invariant, we finally arrive at

$$\eta(X, Y) = \sup \left\{ \rho(Y, \phi(X)) \,\middle|\, \phi \in \mathbb{R}^{\mathcal{X}} \text{ measurable},\ 0 < V(\phi(Y)) < +\infty \right\},$$

showing that the correlation ratio is nothing but a restricted maximum correlation coefficient. In particular, it is possible to use $\text{RMC}(X, Y)_{\phi_{\Theta_{\mathcal{X}}}, \text{Id}_{\mathcal{Y}}, n}$ as an efficient estimator of $\eta(X, Y)$, where $\text{Id}_{\mathcal{Y}}$ denotes the identity function over $\mathcal{Y}$, as we claimed in § 2.

If $\alpha \in \mathbb{R}^k$ achieves the supremum in equation 3.6, then $\sum_{j=1}^k \alpha_j \phi_{\Theta_{\mathcal{X}}^{(j)}}(X)$ is actually (up to a constant) an estimation of $E[Y \mid X]$. In essence, replacing the random nonlinear projections $\phi_{\Theta_{\mathcal{X}}}$ by local polynomials yields the method of Da Veiga et al. (2009) for estimating correlation ratio, which is thus similar to ours but uses different approximation functionals.

## 3.6 Integral Probability Metric and Optimal Transport Cost

In the remaining of this work, we will focus on the three classes of dependence measures presented above. However, we conclude this section by mentioning more general forms of dependence measures, encompassing most of the above and going beyond.

### 3.6.1 Integral Probability Metrics

Considering $P, Q$ two probability distributions over a generic space $\mathcal{Z}$, an integral probability metric between $P$ and $Q$ is defined by Zolotarev (1984) and Müller (1997) as

$$\text{IPM}_{\mathscr{F}}(P, Q) \stackrel{\text{def}}{=} \sup_{\phi \in \mathscr{F}} \left| \int \phi \, dP - \int \phi \, dQ \right|, \tag{3.7}$$

where $\mathscr{F} \subset \mathbb{R}^{\mathcal{Z}}$ is a set of integrable functions. This definition is fairly general, and by appropriately choosing $\mathscr{F}$, one retrieves many popular metrics. Considering the unit ball of the reproducing kernel Hilbert space $\mathcal{H}_k$ of § 3.3.1, $\mathcal{F} \stackrel{\text{set}}{=} \{\phi \in \mathcal{H}_k \mid \|\phi\|_{\mathcal{H}_k} \leq 1\}$, then the difference of integrals in equation 3.7 is the inner product $\langle \phi, \mu_k(P) - \mu_k(Q) \rangle_{\mathcal{H}_k}$ and thus $\text{IPM}_{\mathcal{F}}(P,Q) = \|\mu_k(P) - \mu_k(Q)\|_{\mathcal{H}_k}$ is the kernel distance. Considering the unit ball for the uniform norm, $\mathcal{F} \stackrel{\text{set}}{=} \{\phi \in \mathbb{R}^{\mathcal{Z}} \mid \sup_{z \in \mathcal{Z}} |\phi(z)| \leq 1\}$ it becomes the total variation Csiszár divergence introduced in § 3.4.1; it is however the only nontrivial Csiszár divergence which can be expressed as an integral probability metric, as observed by Sriperumbudur et al. (2012) thanks to a result of Khosravifard et al. (2007). Consider finally the special case of dependence measure, when P and Q are respectively the joint distribution $P_{X,Y}$ and the product of marginals $P_X \otimes P_Y$. In that case $\mathcal{Z} = \mathcal{X} \times \mathcal{Y}$, and if $\phi \in \mathbb{R}^{\mathcal{X} \times \mathcal{Y}}$ is separable as $(x, y) \mapsto \phi_{\mathcal{X}}(x) \phi_{\mathcal{Y}}(y)$ with $\phi_{\mathcal{X}} \in \mathbb{R}^{\mathcal{X}}, \phi_{\mathcal{Y}} \in \mathbb{R}^{\mathcal{Y}}$ measurable with $V(\phi_{\mathcal{X}}(X)) = V(\phi_{\mathcal{Y}}(Y)) = 1$, then the difference of integrals in equation 3.7 becomes $|E(\phi_{\mathcal{X}}(X)\phi_{\mathcal{Y}}(Y)) - E(\phi_{\mathcal{X}}(X))E(\phi_{\mathcal{Y}}(Y))| = |\rho(\phi_{\mathcal{X}}(X), \phi_{\mathcal{Y}}(Y))|$. Setting $\mathcal{F}$ as the set of all functionals satisfying the above, the integral probability metric becomes the maximum correlation coefficient.

Integral probability metrics are thus a convenient unifying tool for studying some of the properties of a wide class of dissimilarity measures; for once, as the name suggests, they all enjoy symmetry and triangle inequality, although they actually are metrics only for $\mathcal{F}$ rich enough. In addition, it is versatile in the sense that it is possible to tailor the set $\mathcal{F}$ according to one's needs; notably for use of dependence measures in conditional and target sensitivity analysis. However, such general form tells very few on the usefulness and estimation of the resulting tool, and one has to study the specificities of each case just as we did above, §§ 3.3–3.5. Alternatively, many interesting integral probability metrics can be studied under a different formulation, which we describe next.

### 3.6.2 Optimal Transport Cost

The theory and application of optimal transport between measures goes well beyond the quantification of dissimilarity between probability distributions. In our setting, that is when P and Q are probability distributions over the same space $\mathcal{Z}$, an optimal transport cost is defined as

$$\text{OTC}_c(P, Q) \stackrel{\text{def}}{=} \inf_{\Pi \in \mathcal{P}(P,Q)} \int_{\mathcal{Z} \times \mathcal{Z}} c(z, z') \, d\Pi(z, z') \, ,$$

where $c: \mathcal{Z} \times \mathcal{Z} \to \mathbb{R}$ is a *cost function*, and $\mathcal{P}(P, Q)$ is the set of all joint probability distributions on $\mathcal{Z} \times \mathcal{Z}$ whose marginals with respect to the first and second variable are P and Q, respectively. This can be thought as the problem of reorganizing "material" initially distributed according to P into a configuration distributed as Q, under the constraint that transporting from $z$ to $z'$ costs $c(z, z')$. Elements of $\mathcal{P}(P, Q)$ are called *transport* (or *transference*) *plans*, and those achieving the infimum are the *optimal transport plans*.

The link with integral probability metrics is established through the powerful *Kantorovich duality theorem* as soon as the cost function is symmetric and satisfies the triangle inequality. In that case, Villani (2008, Theorem 5.10 together with Particular Case 5.4) shows that $\text{OTC}_c = \text{IPM}_{\mathcal{F}}$ with $\mathcal{F} = \{\phi \in \mathbb{R}^{\mathcal{Z}} \mid \forall z, z' \in \mathcal{Z}, \phi(z) - \phi(z') \leq c(z, z')\}$. When $c$ is a metric, the optimal transport cost is called the *Kantorovich distance* (or Wasserstein distance of order 1) and the corresponding set $\mathcal{F}$ is the set of 1-Lipschitz continuous functions. An even more particular case, with the cost $(z, z') \mapsto 0$ if $z = z'$, 1 otherwise, yields the total variation already discussed.

In our opinion, this is in theory the most flexible tool due to the possibility of specifying virtually any cost function, while retaining ease of interpretation. In addition, given any cost function $c$ and samples of the distributions P and Q, an estimator of $\text{OTC}_c(P, Q)$ can be written as the optimal value of a linear program. Let us specify the case of the *transport dependence* *measure* $\text{TDM}_c(X, Y) \overset{\text{def}}{=} \text{OTC}_c(P_X \otimes P_Y, P_{X,Y})$, where the cost is specified between each pair $(x, y), (x', y') \in \mathcal{X} \times \mathcal{Y}$. Then, given $(X^{(i)}, Y^{(i)})_{1 \leq i \leq n}$ independent observations distributed identically to $(X, Y)$, an estimator is

$$\text{TDM}_c(X, Y)_n \overset{\text{def}}{=} \min_{\Pi \in \mathbb{R}^{n^2 \times n}} \sum_{i,j,k=1}^{n} c\big((X^{(i)}, Y^{(j)}), (X^{(k)}, Y^{(k)})\big) \Pi_{(i,j),k}$$

$$\text{subject to} \quad \forall\, i, j, k \in \{1, \ldots, n\}, \Pi_{(i,j),k} \geq 0, \sum_{k'=1}^{n} \Pi_{(i,j),k'} = \frac{1}{n^2}, \text{ and } \sum_{i',j'=1}^{n} \Pi_{(i',j'),k} = \frac{1}{n}.$$

Although it is a linear program, it has dimension $n^3$ and $n^2 + n$ simplex constraints. This is far too large to be useful in any context. As already considered for the estimator of the Csiszár divergence dependence measure in equation 3.3, it is possible to sample the space $\mathcal{X} \times \mathcal{Y}$ with only $\lfloor n/2 \rfloor$ truly independent pairs $(X^{(2i-1)}, Y^{(2i)})_{1 \leq i \leq \lfloor n/2 \rfloor}$ in place of all the $n^2$ possible pairs $(X^{(i)}, Y^{(j)})_{1 \leq i, j \leq n}$, reducing the above linear program to dimension $O(n^2)$ with $O(n)$ simplex constraints. Nevertheless, this remains intractable for most applications, so that transport dependence measure cannot be used under this general form without further drastic improvements, such as acceleration through entropic regularizations as explained by Benamou et al. (2015).

## 4 Some Tools for Conditional and Target Sensitivity Analysis

All of the sensitivity measures detailed above can be easily adapted to target and conditional sensitivity analysis. We describe first general approaches which can be applied to any sensitivity measure. Precisions are then given for each tool that we consider. In addition, we describe some variants, specific to each tool and which do not fall directly into the above general approaches. These variants have their own advantages, and can be seen as hybrid between target and conditional sensitivity analysis, measuring an overall importance of a factor with respect to a critical phenomenon.

### 4.1 Transformations and Weights

Our general approaches are based on transformations of the variable quantifying the phenomenon and on conditioning; specific notions and notations are introduced here.

#### 4.1.1 Targeting with Transformations

In order to study the *occurrences* of the phenomenon $Y$ within the critical domain $\mathcal{C} \subset \mathcal{Y}$, the natural transformation which comes to mind is a binary random variable encoding directly the actual phenomenon of interest and suppressing uninformative fluctuations. This leads to consider the weight function $1_{\mathcal{C}} : \mathcal{Y} \to \{0, 1\} : y \mapsto 1$ if $y \in \mathcal{C}$, 0 otherwise.

Now, recall that a limited number of observations is usually assumed, so that estimation considerations cannot be ignored. The binary transformation above might result in a significant loss of the information conveyed by the relative values of $Y$. Indeed, when the critical probability $P_Y(\mathcal{C})$ is low, most data is summed up to a bunch of zeroes.

Fortunately, a sensible relaxation of the binary assumption can be given as soon as one can evaluate some sort of distance $d_{\mathcal{C}} : \mathcal{Y} \to \mathbb{R}_+$ between each point in $\mathcal{Y}$ and the critical domain $\mathcal{C}$. One can compose it by a decreasing real function $\mathbb{R} \to [0, 1]$, with the rationals that the closer is an observation to the critical domain, the more likely it is to convey similar information; this of course assumes some kind of regularity of the phenomenon's statistical properties. When $\mathcal{Y}$ lies in an Euclidean space, we typically consider the weight function $y \mapsto \exp(-d_{\mathcal{C}}(y)/s)$, where $d_{\mathcal{C}}(y) \stackrel{\text{def}}{=} \inf_{y' \in \mathcal{C}} \|y - y'\|$. Here, the exponential function encodes multiplicative contributions, and $s$ is a smoothing parameter depending typically on a measure of dispersion of the values of $Y$.

In all the following, $w \colon \mathcal{Y} \to [0,1]$ is any kind of the above weight functions, either used deterministically, or as a transformation yielding a random variable through the composition $w(Y)$. Any sensitivity measure between a group of factors $X$ and $w(Y)$ yields a target sensitivity measure.

### 4.1.2 Conditioning with Weighted Probabilities

Alternatively, if one wishes to study the *behaviour* of the phenomenon within the critical domain, a natural idea is *conditioning by the event* $\{Y \in \mathcal{C}\}$. Given an initial probability space $(\Omega, \mathfrak{F}, P)$, if $A \in \mathfrak{F}$ is an event of nonzero probability, then conditioning by $A$ simply means *endowing the measurable space* $(\Omega, \mathfrak{F})$ *with the probability measure* $P_{|A}$, defined as for all $B \in \mathfrak{F}$, $P_{|A}(B) \stackrel{\text{def}}{=} P(B \cap A)/P(A)$. If $X$ is a random variable over $(\Omega, \mathfrak{F}, P)$, then its law conditionally to $A$ is the law of the mapping $X$ over the conditioned probability space $(\Omega, \mathfrak{F}, P_{|A})$, that is $P_{X|A} \stackrel{\text{def}}{=} P_{|A} \circ X^{-1}$.

Just as we introduced smooth relaxation of the binary transformation above, it might be useful to consider extensions of conditioning allowing to take into account some of the information outside the critical domain. This can be easily done by observing that $P_{|A}(B)$ can be expressed as $\int_B 1_A \, dP / \int_\Omega 1_A \, dP$. If $W$ is a positive nonzero random variable over $(\Omega, \mathfrak{F}, P)$ with finite expectation, we define the *probability* $P$ *weighted by* $W$, noted $P^W$, with for all $B \in \mathfrak{F}$, $P^W(B) \stackrel{\text{def}}{=} \int_B W \, dP / \int_\Omega W \, dP$. In other words, $P^W$ is the probability distribution absolutely continuous with respect to $P$ whose density is proportional to $W$. In addition, if $X$ is a generic random variable, we clarify that the notation $P^W_X$ stands for the image measure $(P^W)_X$; although strictly speaking, it cannot be confused with a weighted image measure $(P_X)^W$ since $W$ is defined over $\Omega$ and not over the range of $X$. Let us also exemplify the particular cases of weighted probabilities which are actual conditional probabilities, $P_{|A} = P^{1_A}$, and $P_{X|A} = P^{1_A}_X$.

In a probabilistic framework, any sensitivity measure is defined depending on a (usually implicit) probability space. When conditioning by weight $W$, we change the underlying probability measure, but the mappings defining the random variables are left unchanged; in such case, the notations are prefixed by $[P^W]$. Let us underline here that, provided that the expectations exist, $[P^W] \, E(X) = E(WX)/E(W)$. Another illustration of importance of conditioning with weighted probabilities concerns the estimation of copula transforms. In comparison to § 3.1.2, given $(X^{(i)}, W^{(i)})_{1 \leq i \leq n}$ independent observations distributed identically to $(X, W)$, the empirical distribution function becomes $\frac{1}{\sum_{i=1}^n W^{(i)}} \sum_{i=1}^n W^{(i)} 1_{]-\infty, X^{(i)}]}$.

For conditional sensitivity analysis, we typically use conditioning by weights $W \stackrel{\text{set}}{=} w(Y)$ as defined above.

## 4.2 Correlation Ratio

As presented in § 2, sensitivity indices based on correlation ratio (widely known as Sobol' indices) all consists in (possibly weighted sums of) correlation ratios of the phenomenon $Y$ with well chosen groups of factors, noted generically $X$.

### 4.2.1 Target Correlation Ratio

Correlation ratios can be directly applied to the transformation $w(Y)$, yielding target sensitivity analysis indices based on $\eta^2(X, w(Y))$. Observe that even for multidimensional $Y$, the transformation $w(Y)$ takes values in [0,1], thus sparing us the trouble of interpreting multidimensional extensions of correlation ratio.

### 4.2.2 Conditional Correlation Ratio

Following § 4.1.2, the correlation ratio conditioned by the critical domain is $[P^{w(Y)}] \eta(X, Y)$. It is important to note that even if the factors are independent under P, they usually are not under $P^{w(Y)}$. The covariance estimator in equation 1.1 cannot be used anymore, hindering the estimation of the correlation ratio as explained in § 2. It is thus recommended to estimate it with randomized maximum correlation following § 3.5.3, with $[P^{w(Y)}] \text{RMC}(X, Y)_{\phi_{\Theta_{\mathcal{X}}}, \text{Id}_{\mathcal{Y}}, n}$. In this estimator, the observations are drawn from the actual probability distribution P. The effect of the weighted probability is taken into account by introducing weights when computing the variance-covariance matrices of the transformations of $X$ and $Y$, before applying canonical correlation.

### 4.2.3 Hybrid Correlation Ratio

Alternatively, it is possible to define a conditional correlation ratio by another transformation of $Y$. We have seen that $w(Y)$, keeping no memory of the actual values of $Y$, is more adapted to target sensitivity; for conditional sensitivity, it is preferable to weight multiplicatively the values, as $w(Y)Y$. However, the fact that $w$ vanishes on regions away from the critical domain seems arbitrary: the value zero might not be meaningful for the phenomenon at hand. Since the correlation ratio is a measure of variance, it still seems relevant to set a constant value over these regions, but equal to the expectation of the resulting transformation; they would then not contribute to the variance of the phenomenon. We thus define the transformation

$$Y_w \stackrel{\text{def}}{=} w(Y)Y + (1 - w(Y))y_0 \quad \text{such that} \quad y_0 \stackrel{\text{def}}{=} \text{E}(Y_w) \text{ ; yielding} \quad y_0 = \frac{\text{E}(w(Y)Y)}{\text{E}(w(Y))} .$$

Observe that with $w \stackrel{\text{set}}{=} 1_{\mathcal{C}}$, $\text{E}(w(Y)) = \text{P}(Y \in \mathcal{C})$ and $y_0 = \text{E}[Y \mid Y \in \mathcal{C}]$; more generally, we have $y_0 = [P^{w(Y)}] \text{E}(Y)$. In any case, it is easy to estimate with $\sum_{i=1}^{n} w(Y^{(i)}) Y^{(i)} / \sum_{i=1}^{n} w(Y^{(i)})$, and $\eta^2(X, Y_w)$ can be estimated in turn with any usual method, with the advantage over the conditional correlation ratio in § 4.2.2 that even observations associated to null weight are somehow taken into account.

## 4.3 Kernel Quadratic Dependence Measure

We recall that this dependence measure is also known as Hilbert–Schmidt independence criterion, and is detailed in § 3.3.

### 4.3.1 Target Kernel Quadratic Dependence Measure

Just as with the correlation ratio, target sensitivity measure of a group of factors can be obtained through the weight transformations $w(Y)$, that is to say $\text{QDM}_{k_{\mathcal{X}}, k_{w(\mathcal{Y})}}(X, w(Y))$. Our notation reminds that the kernels depend on the underlying spaces; in the particular case of the binary transformation $w \stackrel{\text{set}}{=} 1_{\mathcal{C}}$, it seems natural to use a categorical kernel for $k_{\{0,1\}}$. Let us mention that this last case was already suggested and briefly illustrated by Da Veiga (2015).

### 4.3.2 Conditional Kernel Quadratic Dependence Measure

The conditional version $[\mathrm{P}^{w(Y)}]$ $\mathrm{QDM}_{k_\mathcal{X},k_\mathcal{Y}}(X,Y)$ is defined through kernel distance as $\left\|\mu_{k_\mathcal{X} k_\mathcal{Y}}\left(\mathrm{P}_{X,Y}^{w(Y)}\right) - \mu_{k_\mathcal{X} k_\mathcal{Y}}\left(\mathrm{P}_X^{w(Y)} \otimes \mathrm{P}_Y^{w(Y)}\right)\right\|^2_{\mathcal{H}_{k_\mathcal{X} k_\mathcal{Y}}}$ and can be again expressed as expectations of kernels analogously to equation 3.1

$$\mathrm{E}\big(k_\mathcal{X}(X,X')k_\mathcal{Y}(Y,Y')\bar{w}(Y)\bar{w}(Y')\big)$$
$$+ \mathrm{E}\big(k_\mathcal{X}(X,X')\bar{w}(Y)\bar{w}(Y')\big)\mathrm{E}\big(k_\mathcal{Y}(Y,Y')\bar{w}(Y)\bar{w}(Y')\big)$$
$$- 2\mathrm{E}\big(k_\mathcal{X}(X,X')k_\mathcal{Y}(Y,Y'')\bar{w}(Y)\bar{w}(Y')\bar{w}(Y'')\big),$$

having taken care of normalizing the weights $\bar{w} \stackrel{\text{def}}{=} \mathrm{E}(w(Y))^{-1} w$. This can also be estimated in an analogous manner, by replacing empirical averages by weighted averages

$$\sum_{i,j=1}^n \left(k_\mathcal{X}\big(X^{(i)},X^{(j)}\big) - \sum_{\ell=1}^n k_\mathcal{X}\big(X^{(i)},X^{(\ell)}\big)\hat{w}\big(Y^{(\ell)}\big)\right)\left(k_\mathcal{Y}\big(Y^{(i)},Y^{(j)}\big) - \sum_{\ell=1}^n k_\mathcal{Y}\big(Y^{(\ell)},Y^{(j)}\big)\hat{w}\big(Y^{(\ell)}\big)\right)$$
$$\times \hat{w}\big(Y^{(i)}\big)\hat{w}\big(Y^{(j)}\big),$$

with empirical normalized weights $\hat{w} \stackrel{\text{def}}{=} \left(\sum_{i=1}^n w\big(Y^{(i)}\big)\right)^{-1} w$.

### 4.3.3 Hybrid Kernel Quadratic Dependence Measure

It is tempting to view the above as a kernel distance over the same probability space but with a different separable kernel over $\mathcal{X} \times \mathcal{Y}$: $\big((x,y),(x',y')\big) \mapsto k_\mathcal{X}(x,x')k_\mathcal{Y}(y,y')w(y)w(y')$. A closer look shows that it is not the case. However, this remains an interesting adaptation, since if $k_\mathcal{Y}: \mathcal{Y}^2 \to \mathbb{R}$ is positive definite and measurable with $\mathrm{E}\big(k_\mathcal{Y}(Y,Y')\big) < +\infty$, then so is

$$k_\mathcal{Y}^w: (y,y') \mapsto k_\mathcal{Y}(y,y')w(y)w(y'),$$

provided that $w: \mathcal{Y} \to \mathbb{R}$ is measurable and bounded. Thus, $\mathrm{QDM}_{k_\mathcal{X},k_\mathcal{Y}^w}(X,Y)$ makes perfect sense and can be seen as a hybrid between target and conditional versions, since the values of $X$ on their own are not affected by the weights. More precisely, the expression in terms of expectations as in equation 3.1 directly gives

$$\mathrm{E}\big(k_\mathcal{X}(X,X')k_\mathcal{Y}(Y,Y')w(Y)w(Y')\big)$$
$$+ \mathrm{E}\big(k_\mathcal{X}(X,X')\big)\mathrm{E}\big(k_\mathcal{Y}(Y,Y')w(Y)w(Y')\big)$$
$$- 2\mathrm{E}\big(k_\mathcal{X}(X,X')k_\mathcal{Y}(Y,Y'')w(Y')w(Y'')\big).$$

showing that $\mathrm{QDM}_{k_\mathcal{X},k_\mathcal{Y}^w}(X,Y) = \mathrm{E}(w(Y))^2 \left\|\mu_{k_\mathcal{X} k_\mathcal{Y}}\left(\mathrm{P}_{X,Y}^{w(Y)}\right) - \mu_{k_\mathcal{X} k_\mathcal{Y}}\left(\mathrm{P}_X \otimes \mathrm{P}_Y^{w(Y)}\right)\right\|^2_{\mathcal{H}_{k_\mathcal{X} k_\mathcal{Y}}}$.

## 4.4 Csiszár Divergence Dependence Measure

We refer to § 3.4 for the definitions of the full and support versions of Csiszár divergence dependence measure.

### 4.4.1 Target Csiszár Divergence Dependence Measure

As previously, target sensitivity measure of a group of factors can be obtained through Csiszár divergence dependence measures with the transformation $w(Y)$, that is $\mathrm{CDM}_\phi(X,w(Y))$ and $\mathrm{sCDM}_\phi(X,w(Y))$. Let us emphasize that in the case of the binary transformation $w \stackrel{\text{set}}{=} 1_\mathcal{C}$, Radon–Nikodym derivatives should be estimated with normalized categorical kernel.

### 4.4.2 Conditional Csiszár Divergence Dependence Measure

The conditional versions are respectively $[\mathrm{P}^{w(Y)}]\ \mathrm{CDM}_\phi(X,Y) = \mathrm{div}_\phi\left(\mathrm{P}^{w(Y)}_{X,Y}, \mathrm{P}^{w(Y)}_X \otimes \mathrm{P}^{w(Y)}_Y\right)$ and $[\mathrm{P}^{w(Y)}]\ \mathrm{sCDM}_\phi(X,Y) = \mathrm{sdiv}_\phi\left(\mathrm{P}^{w(Y)}_X \otimes \mathrm{P}^{w(Y)}_Y, \mathrm{P}^{w(Y)}_{X,Y}\right)$. In the estimators in equations 3.3 and 3.5, the weights are influencing the expectations in each density estimation and each integral, yielding with empirical normalized weights $\hat{w} \stackrel{\text{def}}{=} \left(\sum_{i=1}^n w(Y^{(i)})\right)^{-1} w$,

$$\sum_{i,j=1}^n \phi\left(\frac{\sum_{\ell=1}^n k_{\mathcal{X},\mathcal{Y}}\left((X^{(i)},Y^{(j)}),(X^{(\ell)},Y^{(\ell)})\right)\hat{w}(Y^{(\ell)})}{\left(\sum_{\ell=1}^n k_\mathcal{X}(X^{(i)},X^{(\ell)})\hat{w}(Y^{(\ell)})\right)\left(\sum_{\ell=1}^n k_\mathcal{Y}(Y^{(j)},Y^{(\ell)})\hat{w}(Y^{(\ell)})\right)}\right)\hat{w}(Y^{(i)})\hat{w}(Y^{(j)}),$$

and

$$\sum_{i=1}^n \phi\left(\frac{\left(\sum_{j=1}^n k_\mathcal{X}(X^{(i)},X^{(j)})\hat{w}(Y^{(j)})\right)\left(\sum_{j=1}^n k_\mathcal{Y}(Y^{(i)},Y^{(j)})\hat{w}(Y^{(j)})\right)}{\sum_{j=1}^n k_{\mathcal{X},\mathcal{Y}}\left((X^{(i)},Y^{(i)}),(X^{(j)},Y^{(j)})\hat{w}(Y^{(j)})\right)}\right)\hat{w}(Y^{(i)}),$$

respectively. Versions with nearest-neighbors density estimation can also be easily adapted. For instance, the $k$-th nearest-neighbor distance of the point $(x,y) \in \mathcal{X} \times \mathcal{Y}$ is the smallest distance $d_k$ such that the cumulative sum of the weights of the points within $d_k$ distance to $(x,y)$ reaches $k$. If copula transforms are used, recall that they are also modified by weighted probabilities.

### 4.4.3 Hybrid Csiszár Divergence Dependence Measure

The above conditional version involves the image measure of weighted distributions $\mathrm{P}_X^W$, $\mathrm{P}_Y^W$ and $\mathrm{P}_{X,Y}^W$, but it is not a Csiszár divergence applied to weighted distributions. The latter is however yet another convenient modification.

In the general framework of the Csiszár divergence, recall that P, Q are probability distributions over a space $\mathcal{Z}$; a weight function is thus a measurable function $w_\mathcal{Z} : \mathcal{Z} \mapsto \mathbb{R}_+$ with finite nonzero expectations with respect to P and Q. Interestingly, proposition A.4 and corollary A.1 show that a Csiszár divergence between weighted distributions is nothing but a *weighted Csiszár divergence*; we define accordingly

$$\mathrm{div}_\phi^{w_\mathcal{Z}}(\mathrm{P},\mathrm{Q}) \stackrel{\text{def}}{=} \mathrm{div}_\phi(\mathrm{P}^{w_\mathcal{Z}}, \mathrm{Q}^{w_\mathcal{Z}}), \quad \text{and} \quad \mathrm{sdiv}_\phi^{w_\mathcal{Z}}(\mathrm{P},\mathrm{Q}) \stackrel{\text{def}}{=} \mathrm{sdiv}_\phi(\mathrm{P}^{w_\mathcal{Z}}, \mathrm{Q}^{w_\mathcal{Z}}).$$

As for the dependence measure of interest to us, with $w_\mathcal{Z} : \mathcal{X} \times \mathcal{Y} \to \mathbb{R}^+ : (x,y) \mapsto w(y)$, notice that $\int w_\mathcal{Z}\, \mathrm{d}(\mathrm{P}_X \otimes \mathrm{P}_Y) = \int w_\mathcal{Z}\, \mathrm{d}\mathrm{P}_{X,Y} = \mathrm{E}(w(Y))$. Finally, propositions A.5 and A.6 show further that

$$\mathrm{CDM}_\phi^{w_\mathcal{Z}}(X,Y) \stackrel{\text{def}}{=} \mathrm{div}_\phi^{w_\mathcal{Z}}(\mathrm{P}_{X,Y}, \mathrm{P}_X \otimes \mathrm{P}_Y) = \mathrm{div}_\phi\left(\mathrm{P}_{X,Y}^{w(Y)}, \mathrm{P}_X \otimes \mathrm{P}_Y^{w(Y)}\right),$$

and

$$\mathrm{sCDM}_\phi^{w_\mathcal{Z}}(X,Y) \stackrel{\text{def}}{=} \mathrm{sdiv}_\phi^{w_\mathcal{Z}}(\mathrm{P}_X \otimes \mathrm{P}_Y, \mathrm{P}_{X,Y}) = \mathrm{sdiv}_\phi\left(\mathrm{P}_X \otimes \mathrm{P}_Y^{w(Y)}, \mathrm{P}_{X,Y}^{w(Y)}\right),$$

establishing the link with the conditional Csiszár divergence dependence measure.

### 4.5 Randomized Maximum Correlation

Recall from § 3.5.2 that randomized maximum correlation is defined through a set of nonlinear functionals applied to the random variables.

### 4.5.1 Target Randomized Maximum Correlation is a Correlation Ratio

So far, our approach for target sensitivity analysis is precisely to work on a specific real transformation of the studied phenomenon $w(Y)$. In this context, there seems to be no interest in specifying additional transformations on it. The target randomized maximum correlation ends up being of the form $\text{RMC}(X, Y)_{\phi_{\Theta_{\mathcal{X}}}, w, n}$ or equivalently $\text{RMC}(X, w(Y))_{\phi_{\Theta_{\mathcal{X}}}, \text{Id}_{[0,1]}, n}$. But, as discussed in § 3.5.3, this is nothing but an efficient estimator of the target correlation ratio $\eta(X, w(Y))$.

Let us add that it is possible to adapt the set of nonlinear functionals $\phi_{\Theta_{\mathcal{X}}}$ to the modeling of weights functions such as the ones we specified in § 4.1.1. We propose in particular *logistic functions* over Euclidean spaces $\mathcal{X} \subseteq \mathbb{R}^p$, $\phi_{\theta,b} : \theta \mapsto \left(1 + \exp(-\langle \theta, x \rangle + b)\right)^{-1}$, with parameters $\theta \in \mathbb{R}^p$ and $b \in \mathbb{R}$ drawn according to normal distributions scaled by the dimension $p$.

### 4.5.2 Conditional Randomized Maximum Correlation

In contrast to the above, when conditioning by a weight $W$, it still makes perfect sense to consider any nonlinear functionals of the phenomenon. Thus, the conditional randomized maximum correlation $\left[\text{P}^{w(Y)}\right] \text{RMC}(X, Y)_{\phi_{\Theta_{\mathcal{X}}}, \phi_{\Theta_{\mathcal{Y}}}, n}$ is different, and able to capture more complex dependence, than its correlation ratio counterpart discussed in § 4.2.2. We underline once again that the effect of the weighted probability is taken into account by introducing weights when computing the variance-covariance matrices of the transformations of $X$ and $Y$, before applying canonical correlation; and that if copula transforms are used, they are also modified by weighted probabilities.

## 5 Numerical Illustrations

We conduct here numerical illustrations of the different tools we presented so far. First, we show that dependence measures are interesting for global sensitivity analysis. Then, we focus on the particular case of estimating correlation ratio with randomized maximum correlation approach. Finally, we demonstrate on concise examples that target and conditional sensitivity analysis explore aspects of a model which are both different from global sensitivity analysis and valuable for practitioners.

Note that all the above tools are implemented in the language R, interfaced with C++ for some routines; we intend to integrate them to the *Sensitivity* package of R.

### 5.1 Global Sensitivity Analysis with Dependence Measures

In order to illustrate the behavior of dependence measures, we use two synthetic models, well-known from the sensitivity analysis community: the so-called Sobol' $g$ function, and the Ishigami–Homma function. Their main properties are recalled below.

### 5.1.1 Experiments with the Sobol' $g$ model

This model is specifically designed to study sensitivity analysis through correlation ratios (the Sobol' indices). It is defined in dimension $d \in \mathbb{N} \setminus \{0\}$ by

$$f : x \mapsto \prod_{i=1}^{d} g_i(x_i), \quad \text{with for all } i \in \{1, \ldots, d\}, \quad g_i : x_i \mapsto \frac{|4x_i - 2| + a_i}{1 + a_i},$$

where $a_i \in \mathbb{R}_+$; all factors $(X_1, \ldots, X_d)$ are independent and uniformly distributed over $[0,1]$. Each function $g_i$ is conveniently nonlinear, nonmonotonous, nondifferentiable, satisfying $\int_0^1 g_i(x_i)\,dx_i = 1$ and $\int_0^1 g_i(x_i)^2\,dx_i = 1 + \frac{1}{3(1+a_i)^2}$. Thus, $E(Y) = 1$, $V(Y) = \prod_{i=1}^d \left(1 + \frac{1}{3(1+a_i)^2}\right) - 1$, and for all $i \in \{1, \ldots, d\}$, the first-order and total-order Sobol' indices yield respectively

$$\eta^2(X_i, Y) = \frac{1}{3(1+a_i)^2 V(Y)} \quad \text{and} \quad 1 - \eta^2(X_{c_{\{i\}}}, Y) = 1 - \frac{\prod_{j \neq i}\left(1 + \frac{1}{3(1+a_j)^2}\right) - 1}{V(Y)}.$$

In consequence, the higher the $a_i$, the less influential the $X_i$; although the correlation ratios seem the most suitable to put this in evidence, we propose to test some dependence measures with rankings of the factors in mind. We set $d \stackrel{\text{set}}{=} 4$ and the 4-tuple parameter $a \stackrel{\text{set}}{=} (0, 1, 9, 99)$.

We consider the sensitivity measures described in table 1; note in particular that we test "total-order" versions of the dependence measures, in analogy to the total-order Sobol' indices. This illustrates notably the possibility of considering multidimensional factors. The principle is to apply a decreasing function to dependence measures of the group of factors complementary to the one investigated. Here, we use the complement to unity, which makes sense because we normalize the dependence measures; but there is no further justification, in contrast to the Sobol' indices which are based on ratios of explained variance.

Table 1: Sensitivity measures used on the Sobol' $g$ model. In analogy to the total-order correlation ratio sensitivity measure, we define "total-order" versions of the dependence measures.

| Notation | Definition | Expression for factor $i$ |
| --- | --- | --- |
| $S_{PF}^{(1)}$ $S_{PF}^{(tot)}$ | Correlation ratio sensitivity measure (Sobol' indices), estimated with pick-and-freeze factors combinations as in equation 2.1 | $\eta^2(X_i, Y)_n$ $1 - \eta^2(X_{c_{\{i\}}}, Y)$ |
| $QDM_G^{(1)}$ $QDM_G^{(tot)}$ | Normalized kernel quadratic dependence measure with Gaussian kernels with scalar bandwidth,[a] determined by the median of the observed distances | $\overline{QDM}_{k_{\mathcal{X}}, k_{\mathcal{Y}}}(X_i, Y)_n$ $1 - \overline{QDM}_{k_{\mathcal{X}}, k_{\mathcal{Y}}}(X_{c_{\{i\}}}, Y)_n$ |
| $MI_G^{(1)}$ $MI_G^{(tot)}$ | Normalized mutual information, Gaussian kernel density estimation with diagonal bandwidth following Silverman (1986) rule-of-thumb | $\overline{sCDM}_{-\log}(X_i, Y)_{k_G, n}$ $1 - \overline{sCDM}_{-\log}(X_{c_{\{i\}}}, Y)_{k_G, n}$ |
| $RMC_{c,G,s}^{(1)}$ $RMC_{c,G,s}^{(tot)}$ | Square randomized maximum correlation as defined by Lopez-Paz et al. (2013) (copula transforms, Gaussian weights and intercept, sine nonlinearity), and $\sqrt{n}$ nonlinear projections for both $X$ and $Y$ | $RMC(X_i, Y)^2_{\phi_{\Theta_{\mathcal{X}}}, \phi_{\Theta_{\mathcal{Y}}}, n}$ $1 - RMC(X_{c_{\{i\}}}, Y)^2_{\phi_{\Theta_{\mathcal{X}}}, \phi_{\Theta_{\mathcal{Y}}}, n}$ |

[a] A scalar bandwidth cannot account for factors with different orders of magnitude, but here the factors are identically distributed

All sensitivity measures are computed on samples of size $n \stackrel{\text{set}}{=} 50, 200$, and $1\,000$. For the pick-and-freeze estimator of the correlation ratio, the sample is divided in two equal parts so that the model is computed on $n$ factors combinations following equation 2.1. Note that this requires $\frac{n}{2} \times (d+1)$ specific model computations for all first-order indices, plus an additional $\frac{n}{2}$ for all total-order indices. In contrast, all the dependence measures are computed from the same pseudo-random sample of size $n$. This fact should be kept in mind by the practitioner when evaluating the relative relevance of the sensitivity measures.

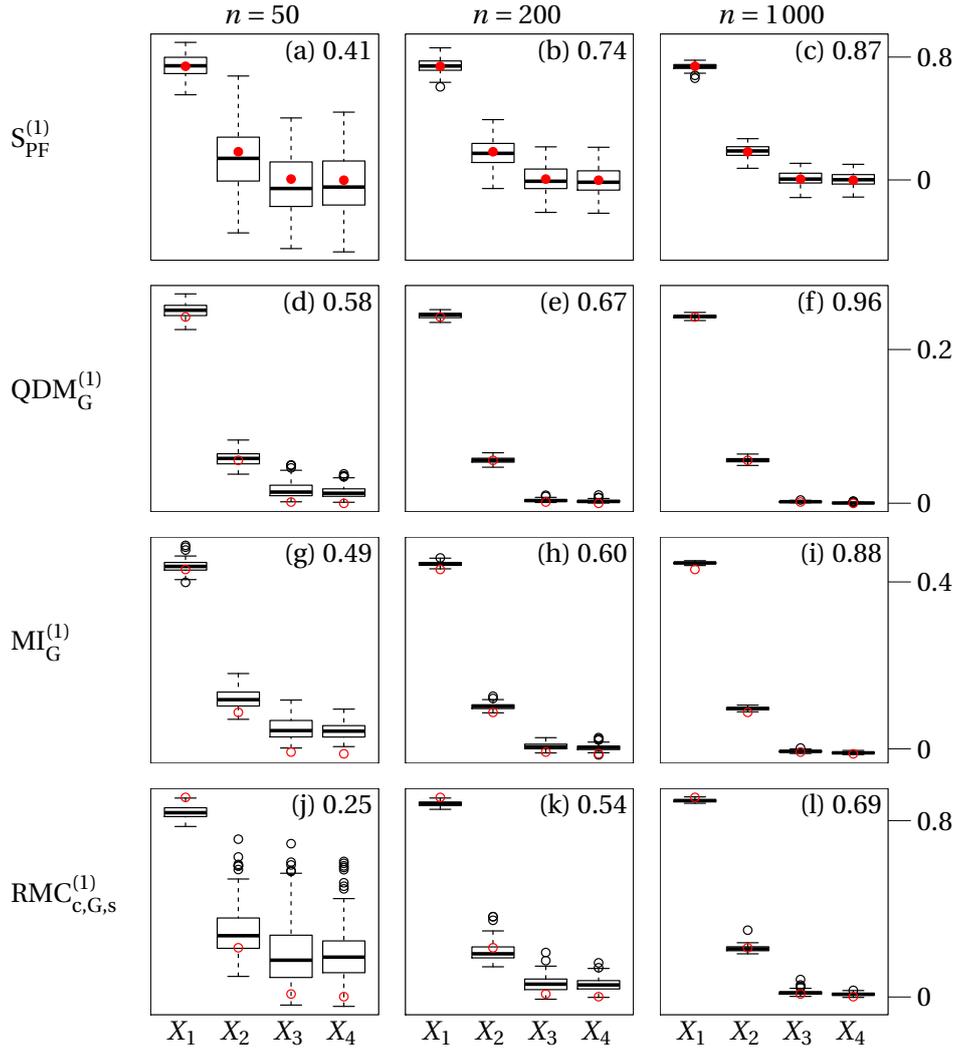

Figure 2: First-order global sensitivity measures on the Sobol' $g$ model. Filled red dots are analytical values, hollow red circles are asymptotic values estimated with sample sizes $n \stackrel{\text{set}}{=} 10\,000$. The number given in the top left corner of each plot is the proportion of repetitions with correct ordering.

Since all considered sensitivity measures rely at some point on Monte Carlo estimations of expectations, we draw the samples with *low-discrepancy space-filling methods* (see for instance the algorithms of Damblin et al., 2013). For each sample size, we draw hundred repetitions, and schematize the resulting distributions of the sensitivity measures with Tuckey box plots. In addition, the analytical values of the Sobol' indices are reported as filled red dots. For all other measures, we report with hollow red circles asymptotic values, estimated as the median over ten repetitions with sample size $n \stackrel{\text{set}}{=} 10\,000$. Since we know that the factors should be ranked in descending order, we also report the proportions of repetitions for which the factors were correctly ordered.

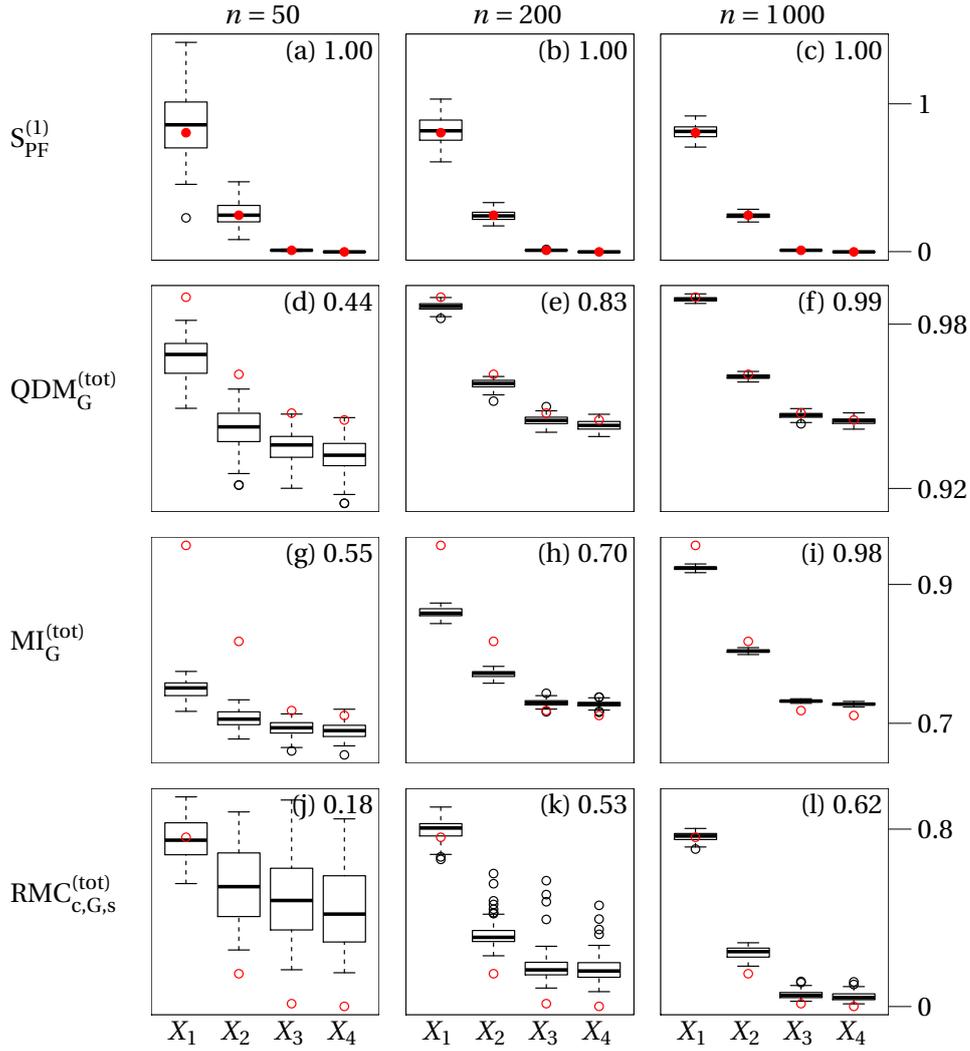

Figure 3: Total-order global sensitivity measures on the Sobol' *g* model. See figure 2.

The general ranking can already be seen from all first-order sensitivity measures on figure 2. They all have at most a small bias, except perhaps the randomized maximum correlation and the mutual information at very low sample size. However, the quadratic dependence measure and the mutual information have much less variability across repetitions than the two others, as can be seen by the relative heights of the box plots. Surprisingly, this better precision does not necessarily translate into better ranking ability. Consider that the correlation ratio computed with pick-and-freeze factors combinations performs as well as the other two in that respect. It seems to enjoy the property that, on a given sample, it either overestimates all indices together, or underestimates them all together, hence preserving the relative ordering. Randomized maximum correlation does not enjoy such property, recording less correct ordering in spite of more precision for sample sizes larger than $n \stackrel{\text{set}}{=} 200$.

Total-order versions exhibits similar behavior on figure 3. Multidimensional versions of the quadratic dependence measure and mutual information have an increased bias relatively to the differences between factors, but precision and ordering is preserved, and even enhanced for sample sizes larger than $n \stackrel{\text{set}}{=} 200$. They can thus be used with multidimensional factors, at least when the dimension is not too large. Randomized maximum correlation seems less robust to the increase of the dimension. The favorable behavior of pick-and-freeze estimator of correlation ratio for factors ranking is even more opportune with the total-order indices. On figure 3(a), it is able to rank correctly all repetitions with sample size as little as $n \stackrel{\text{set}}{=} 50$, in spite of large variability, especially on the factor $X_1$. This deserves to be noted, since in this example with only four factors, the total number of required model computations is only $25 \times 5 = 125$. This must however be qualified by the fact that the Sobol' $g$ model is simplistic and specifically adapted to analysis with correlation ratios. Moreover, consider that the ranges of observed dependence measures of the first three factors are almost nonintersecting, on each of the figures 2(e), 2(h), 3(e) and 3(h). Thus, the majority of rankings errors made by the quadratic dependence measure and mutual information concern the ordering of the last two factors, which in most applications would both be deemed as "noninfluential" anyway.

### 5.1.2 Experiments with the Ishigami–Homma model

This model is also designed to study sensitivity analysis through correlation ratios, but in a somewhat more challenging way. It is defined in dimension $d \stackrel{\text{set}}{=} 3$ by

$$f\colon x \mapsto \sin(x_1) + a\sin^2(x_2) + bx_3^4 \sin(x_1),$$

where $a, b \in \mathbb{R}_+$; all factors $(X_1, X_2, X_3)$ are independent and uniformly distributed over $[-\pi, \pi]$. The influence of the factor $X_2$ is purely additive, its importance being modulated by the parameter $a$. The influence of the factor $X_1$ has an additive part and an interaction with the factor $X_3$, the balance being tuned by parameter $b$.

Using the following integrals,

$$\int_{-\pi}^{\pi} \sin(x)\,\mathrm{d}x = 0, \quad \frac{1}{2\pi}\int_{-\pi}^{\pi}\sin(x)^2\,\mathrm{d}x = \frac{1}{2}, \quad \frac{1}{2\pi}\int_{-\pi}^{\pi}\left(1+bx^4\right)\mathrm{d}x = 1 + b\frac{\pi^4}{5},$$

and then

$$\frac{1}{2\pi}\int_{-\pi}^{\pi}\sin(x)^4\,\mathrm{d}x = \frac{3}{8}, \quad \frac{1}{2\pi}\int_{-\pi}^{\pi}\left(1+bx^4\right)^2\mathrm{d}x = \left(1+b\frac{\pi^4}{5}\right)^2 + \frac{16\pi^8}{225},$$

one gets $\mathrm{E}(Y) = \frac{a}{2}$, $\mathrm{V}(Y) = A + B + C$ where $A \stackrel{\text{def}}{=} \frac{a^2}{8}$, $B \stackrel{\text{def}}{=} \frac{1}{2}\left(1 + b\frac{\pi^4}{5}\right)^2$ and $C \stackrel{\text{def}}{=} b^2\frac{8\pi^8}{225}$, and

$$\eta^2(X_1, Y) = \frac{B}{\mathrm{V}(Y)}, \qquad \eta^2(X_2, Y) = \frac{A}{\mathrm{V}(Y)}, \qquad \eta^2(X_3, Y) = 0,$$

$$1 - \eta^2((X_2, X_3), Y) = \frac{B+C}{\mathrm{V}(Y)}, \quad 1 - \eta^2((X_1, X_3), Y) = \frac{A}{\mathrm{V}(Y)}, \quad 1 - \eta^2((X_1, X_2), Y) = \frac{C}{\mathrm{V}(Y)}.$$

We set the parameters $a \stackrel{\text{set}}{=} 5$ and $b \stackrel{\text{set}}{=} 0.1$, so that $\eta^2(X_1, Y) = 0.40$, $\eta^2(X_2, Y) = 0.29$ and $\eta^2(X_3, Y) = 0$, while $1 - \eta^2(X^c_{\{1\}}, Y) = 0.71$, $1 - \eta^2(X^c_{\{2\}}, Y) = 0.29$, and $1 - \eta^2(X^c_{\{3\}}, Y) = 0.31$. In particular, the correlation ratio between $X_3$ and $Y$ is zero, but the total contribution of $X_3$ in the variance of $Y$ is actually higher than the one of $X_2$.

In addition to the sensitivity measures described in table 1, we also consider some modifications given in table 2. We compute hundred repetitions of space-filling samples of size $n \stackrel{\text{set}}{=} 1\,000$, and schematize the resulting distributions of the sensitivity measures with Tuckey box plots on figure 4; again, we report analytical values with filled red dots and estimated asymptotic values with hollow red circles.

The global sensitivity analysis of the Ishigami–Homma model is harder to interpret than for the Sobol' $g$ model. As expected, the first-order and the total-order correlation ratio give different factors rankings; moreover the pick-and-freeze estimation has low precision, especially for identifying the zero correlation ratio between $X_3$ and $Y$ in figure 4(a). In contrast to the correlation ratio, first-order dependence measures are able to identify the importance of factor $X_3$: the measures are clearly nonzero on figures 4(b)–(f).

Table 2: Additional dependence measures used on the Ishigami–Homma model.

| Notation | Definition | Expression for factor $i$ |
|---|---|---|
| $\text{QDM}_{c,G}^{(1)}$ $\text{QDM}_{c,G}^{(\text{tot})}$ | Normalized kernel quadratic dependence measure on the copulas with Gaussian kernel | $\overline{\text{QDM}}_{k_G}(F_X(X)_i, F_Y(Y))_n$ $1 - \overline{\text{QDM}}_{k_G}(F_X(X)_{c\{i\}}, F_Y(Y))_n$ |
| $\text{MI}_{c,\text{nn}}^{(1)}$ $\text{MI}_{c,\text{nn}}^{(\text{tot})}$ | Normalized mutual information with truncated nearest-neighbors copula density estimation, number of neighbors set as $n^{4/5}$ (see Silverman, 1986, p. 99) | $\overline{\text{sCDM}}_{-\log}(X_i, Y)_{k_{\text{nn}}, n}$ $1 - \overline{\text{sCDM}}_{-\log}(X_{c\{i\}}, Y)_{k_{\text{nn}}, n}$ |

However, some weaknesses of the dependence measures are highlighted by this experiment. First, the randomized maximum correlation is severely imprecise in figure 4(d). Then, the quadratic dependence measure, both with and without prior copula transforms, seems to consider $X_2$ as noninfluential in figures 4(b) and (c); although this should be further investigated through statistical tests of independence (typically, by permutation). Also, all occurrences of mutual information suffers from significant bias in figures 4(e), (f), (k), and (l). Fortunately, this bias does not affect the order of magnitude nor the ordering of the measures along factors.

A last concern is that almost all dependence measures considered here attribute more relative importance to the factor $X_2$ with their total-order versions than with their first-order versions (compare figures 4(b)–(e) with their corresponding figures 4(h)–(k)); sometimes to the point of inverting the importance rank with $X_3$. This seems wrong to us, since the contribution of $X_2$ is only additive, in contrast to $X_1$ and $X_3$. A closer look highlights what could be the problem. For instance, all quadratic dependence measures for the first factor, $\overline{\text{QDM}}_{k_{[-\pi,\pi]}, k_{\mathcal{Y}}}(X_1, Y)_n$, are over 0.2 in figure 4(b). In the same time, the total-order versions for the second factor $1 - \overline{\text{QDM}}_{k_{[-\pi,\pi]^2}, k_{\mathcal{Y}}}((X_1, X_3), Y)_n$ are also all over 0.8 in figure 4(h). This means that, on all repetitions, the measure of the dependence between $X_1$ and $Y$ is greater than the measure of the dependence between $(X_1, X_3)$ and $Y$. Although this is not the desired behavior, observe that $X_1$ and $(X_1, X_3)$ obviously lie in different spaces, over which quadratic dependence measures are defined with different kernels and are thus not necessarily comparable. The mutual information seems somewhat less affected in that respect, but requires larger samples for accuracy. Recalling § 3.2.1, this confirms that comparing factors of different natures is not straightforward, and that better normalization schemes are yet to be found.

## 5.2 Estimating Correlation Ratio with Randomized Maximum Correlation

We explore here numerically the use of the randomized maximum correlation approach for computing correlation ratios, following § 3.5.3. More precisely, we note respectively $S_{\text{RMC}}^{(1)}$ and $S_{\text{RMC}}^{(\text{tot})}$ the randomized maximum correlation estimators as described in table 1, except that the nonlinear projections of $Y$ are reduced to the identity function; resulting expressions for factor $i$ are respectively $\text{RMC}(X_i, Y)^2_{\phi_{\Theta_{\mathcal{X}}}, \text{Id}_{\mathcal{Y}}, n}$ and $1 - \text{RMC}(X_{c\{i\}}, Y)^2_{\phi_{\Theta_{\mathcal{X}}}, \text{Id}_{\mathcal{Y}}, n}$. Debiasing as explained in § 3.5.2 is also of importance here.

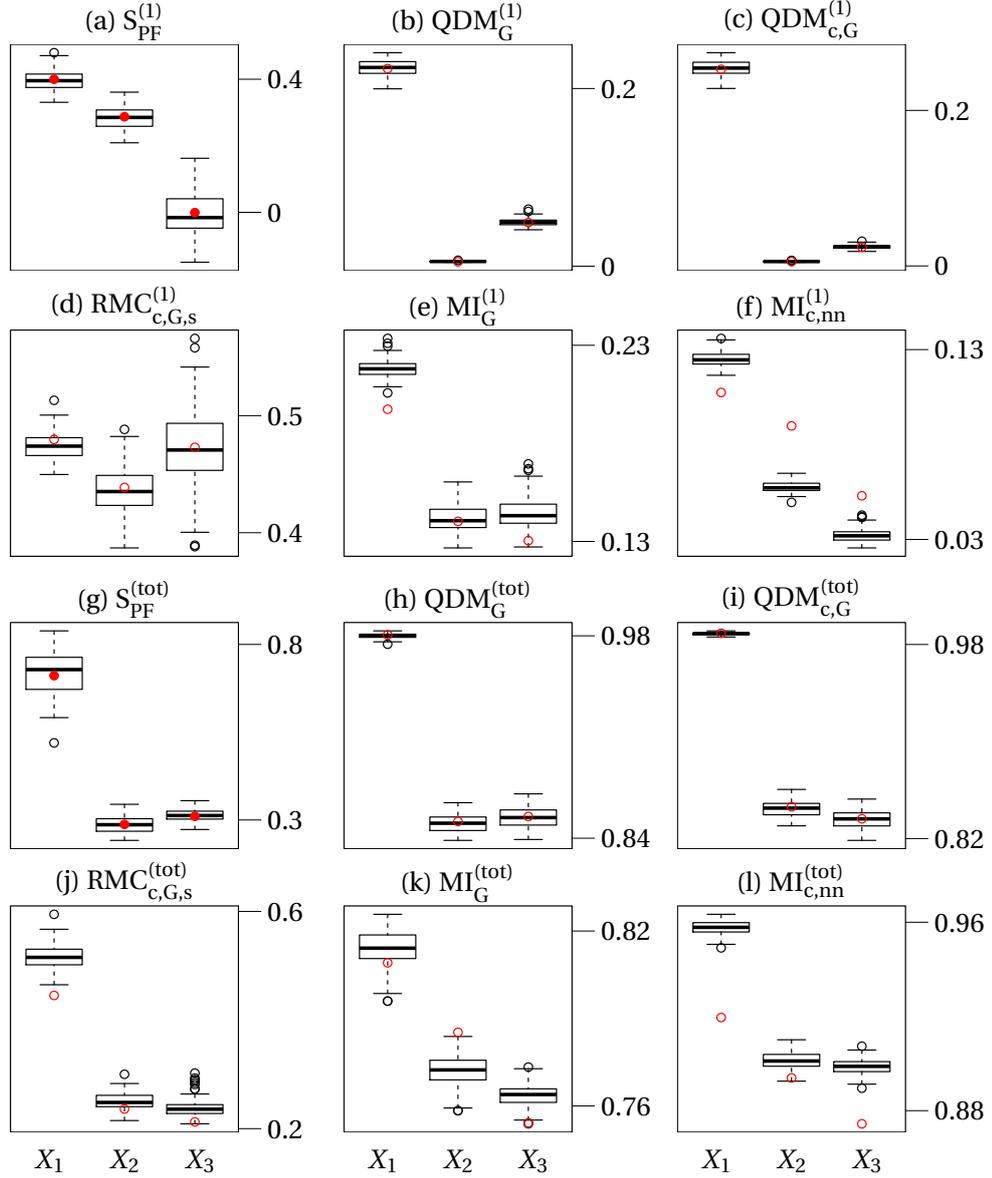

Figure 4: Global sensitivity measures on the Ishigami–Homma model, sample size $n \stackrel{\text{set}}{=} 1\,000$. Filled red dots are analytical values, hollow red circles are asymptotic values estimated with sample sizes $n \stackrel{\text{set}}{=} 10\,000$. Two first rows (a)–(f) show first-order versions, two last rows (g)–(l) show total-order versions.

We compare it to the classical pick-and-freeze estimators on the Sobol' $g$ and the Ishigami–Homma models, whose description, together with analytical values of first-order and total-order correlation ratios, have been given above § 5.1. We compute hundred repetitions of space-filling samples of size $n \stackrel{\text{set}}{=} 200$ and $1\,000$, and schematize the resulting distributions of the estimated correlation ratios with Tuckey box plots on figures 5 and 6, together with the analytical values indicated by red dots.

Recall that the pick-and-freeze estimator requires $\frac{n}{2} \times (d+1)$ well chosen model computations for obtaining all indices of a given order. Thus, if there were nine factors to study, pick-and-freeze estimates with samples of size $n \stackrel{\text{set}}{=} 200$, on the leftmost plots labeled (a) and (e), should be rather compared to randomized maximum correlation estimates with samples of size $n \stackrel{\text{set}}{=} 1\,000$, on the rightmost plots labeled (d) and (h).

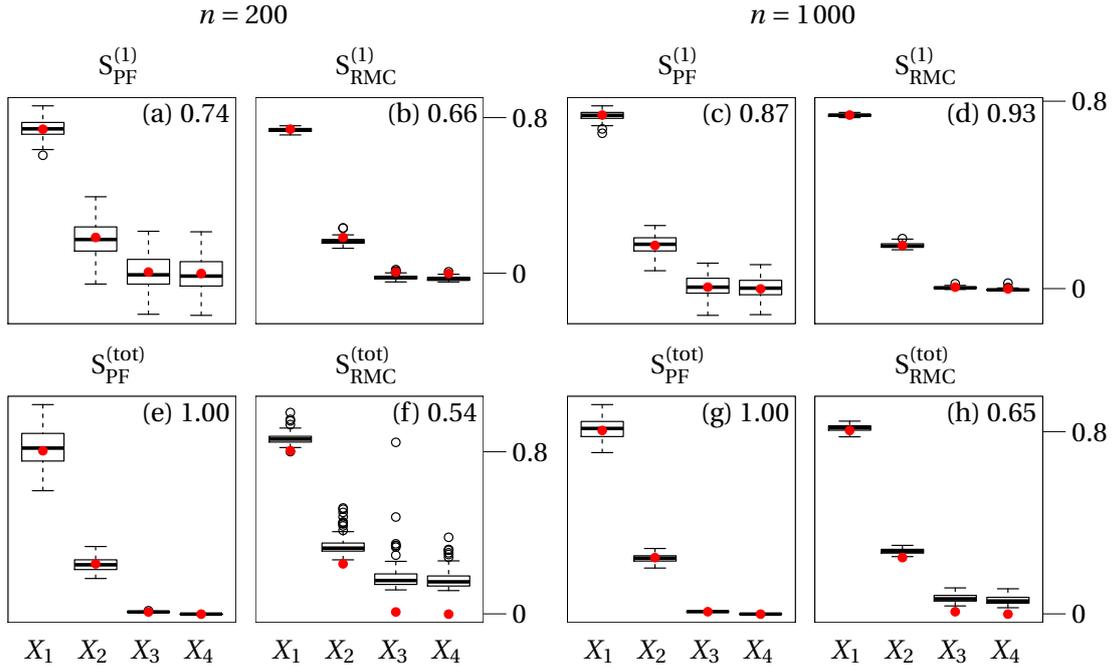

Figure 5: Estimation of correlation ratio sensitivity measures on the Sobol' $g$ model. Analytical values are indicated by red dots. The number given in the top left corner of each plot is the proportion of repetitions with correct ordering of the factors. First row (a)–(d) shows first-order versions, second row (e)–(h) shows total-order versions.

In spite of this, it appears clearly that the randomized maximum correlation approach is more precise for all first-order indices for both models and both sample sizes. As already discussed in § 5.1.1, this better precision does not always translate into better factors ranking on the Sobol' $g$ model, although one can underline the surprisingly good accuracy on figure 5(h).

The performance for the total-order indices are not as good as for the first-order indices. For high indices (factor $X_1$ in both models), the variability of the randomized maximum correlation estimator is still clearly lower than for the pick-and-freeze estimator. This is not true anymore for low indices (factors $X_3$ and $X_4$ for Sobol' $g$, $X_2$ and $X_3$ for Ishigami–Homma). More importantly, the randomized maximum correlation estimator suffers from an overestimation bias (that is, the multidimensional correlation ratios are underestimated), and the occurrence of severe outliers. These flaws decrease rapidly with the sample size, and we believe that randomized maximum correlation remains a good approach for the estimation of multidimensional correlation ratio, at least when the dimension is not too large.

Moreover, we believe that there remains room for better debiasing and other improvements of the randomized maximum correlation. So far, its use for sensitivity analysis seems more successful with its correlation ratio version $S_{\text{RMC}}^{(1)}$ than with its full maximum correlation version $\text{RMC}_{c,G,s}^{(1)}$, as can be seen by comparing figures 5 and 6 with figures 2–4. In its current version, it already constitutes a good candidate for replacement of the ubiquitous pick-and-freeze estimator, requiring no specific experimental design, smaller sample sizes, and no independence assumption. It should also be interesting to compare it with other alternative estimators of correlation ratios as mentioned in § 2, in particular the method of Da Veiga et al. (2009).

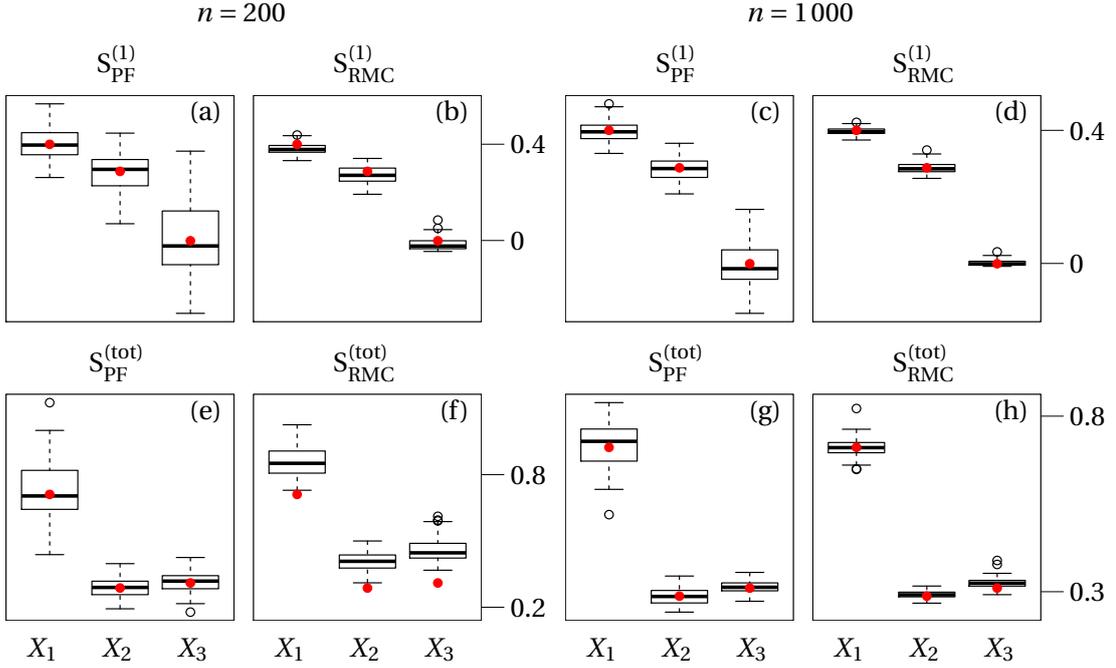

Figure 6: Estimation of correlation ratio sensitivity measures on the Ishigami–Homma model. Analytical values are indicated by red dots. First row (a)–(d) shows first-order versions, second row (e)–(h) shows total-order versions.

### 5.3 Target and Conditional Sensitivity Analysis

We now turn to the illustration of target and conditional sensitivity analysis. To this end, we first propose a model with a simple but strong nonlinearity, which we call *minimum normal uniform*. It is defined in dimension $d \stackrel{\text{set}}{=} 2$, with $f \colon x \mapsto \min(x_1, x_2)$, with independent factors conveniently noted $X_1 \stackrel{\text{set}}{=} N$ and $X_2 \stackrel{\text{set}}{=} U$, following respectively a standard normal distribution, and an uniform distribution over [0,1]. We then explore the more complicated Ishigami–Homma model presented above.

In both models, we suppose that the critical domain is defined by $Y$ exceeding a given critical value $\mathscr{C} \stackrel{\text{set}}{=} \{y \in \mathscr{Y} \mid y \geq c\}$, chosen as the *ninth decile* of $Y$ computed empirically, $c \stackrel{\text{set}}{=} F_{Y,n}^{-1}(0.9)$. Recall that target and conditional sensitivity measures are defined via weight functions $w \colon \mathscr{Y} \to [0,1]$ which depends on $\mathscr{C}$. In both models, we use the indicator $\mathbb{1}_\mathscr{C}$, and a smooth relaxation in accordance with the notion of distance over the reals,

$$w_\mathscr{C} \colon y \mapsto \exp\left(-\frac{\max(c-y,0)}{s\,\sigma_Y}\right); \tag{5.1}$$

where $\sigma_Y$ is an estimation of the standard deviation of $Y$, and $s \stackrel{\text{set}}{=} 1/5$ is a factor tuning the smoothness, chosen so that $w_\mathscr{C}$ almost vanishes one standard deviation away from $\mathscr{C}$.

Among the large choice of interesting sensitivity measures, we consider those in table 3. Correlation ratios estimated with pick-and-freeze factors combinations are included because they are currently the most popular for global sensitivity analysis. Recall however from §§ 4.2.2 and 4.2.3 that they do not allow for proper conditional versions, because conditioning introduces dependence between factors; we use then what we call the "hybrid" version. We report here results only for the first-order indices, but we can mention that the total-order indices behave similarly for target and conditional sensitivity analysis of our models.

Then, we include the quadratic dependence measure with Gaussian kernel, and the mutual information dependence measure with truncated nearest-neighbors copula density estimation, because they appeared to be the most robust for global sensitivity analysis in § 5.1.2. We consider only target and conditional versions, and leave the exploration of their "hybrid" versions to a future work. For the binary target versions, recall that $1_\mathscr{C}(Y)$ is a discrete random variable over $\{0,1\}$. For the mutual information, its law is estimated by empirical frequencies and the law of the joint $(X_i, 1_\mathscr{C}(Y))$ is estimated by conditioning. For the quadratic dependence measure, we use a categorical kernel for $k_{\{0,1\}}$.

Table 3: Sensitivity measures used for target and conditional analysis experiments. The generic weight function $w$ is either $1_\mathscr{C}$, or the smooth relaxation $w_\mathscr{C}$ defined in equation 5.1.

| Notation | Definition | Expression for factor $i$ |
| --- | --- | --- |
| $S_{PF}^{(1,\text{tgt},w)}$ | First-order correlation ratio target sensitivity measure | $\eta^2(X_i, w(Y))_n$ |
| $S_{PF}^{(1,\text{hbd},w)}$ | First-order correlation ratio hybrid sensitivity measure | $\eta^2(X_i, Y_w)_n$ |
| $\text{QDM}_G^{(\text{tgt},w)}$ | Normalized target kernel quadratic dependence measure | $\overline{\text{QDM}}_{k_\mathscr{X}, k_{w(\mathscr{Y})}}(X_i, w(Y))_n$ |
| $\text{QDM}_G^{(\text{cnd},w)}$ | Normalized conditional kernel quadratic dependence measure | $[\text{P}^{w(Y)}]\overline{\text{QDM}}_{k_\mathscr{X}, k_\mathscr{Y}}(X_i, Y)_n$ |
| $\text{MI}_{c,nn}^{(\text{tgt},w)}$ | Normalized target mutual information | $\overline{\text{sCDM}}_{-\log}(X_i, w(Y))_{k_{nn}, n}$ |
| $\text{MI}_{c,nn}^{(\text{cnd},w)}$ | Normalized conditional mutual information | $[\text{P}^{w(Y)}]\overline{\text{sCDM}}_{-\log}(X_i, Y)_{k_{nn}, n}$ |

For each model, we draw hundred different samples of size $n \stackrel{\text{set}}{=} 1\,000$ and schematize the resulting distribution of each conditional or target sensitivity measure, together with their global sensitivity counterpart, with Tuckey box plots on figures 7 and 8.

On the minimum normal uniform model, the critical value is $c = 0.62$. The global analysis, on figures 7(a)–(c), is unanimous: the factor $N$ is much more important than the factor $U$. This is not surprising, since $N$ presents more variability and takes values far below the minimum of $U$.

The target analysis indicates that the ordering of the factors is the same, although the relative importance difference is less drastic. This is again not surprising because $N$ has a higher probability to be below the critical value than $U$, hence still determining again the outcome of interest here, but in the same time, the variability of $N$ below the threshold has no influence anymore. The correlation ratio on figure 7(d), is much less precise than the the dependence measures (e) and (g). The target mutual information shows an important bias, but once again this does not impact the ordering of the factors. It can be noted that the smoothed versions present less variability while still ordering correctly the factors. However, it is unclear if this is thanks to better behavior of the smooth estimator, or simply because the estimated smoothed quantity is some kind of interpolation between target and global measures. In the latter case, this effect would turn out unfavorable if the ordering of the factors were different in both analysis. The smoothed target mutual information on figure 7(i) is clearly problematic, as it yields the same importance measures as the global version (c). This can be explained by the fact that the density estimation is based on copula transforms, and that $Y$ and $w_{\mathscr{C}}(Y)$ have very similar copula transforms with the level of smoothing that we used; in this case and for this particular estimator, smoothing is not judicious.

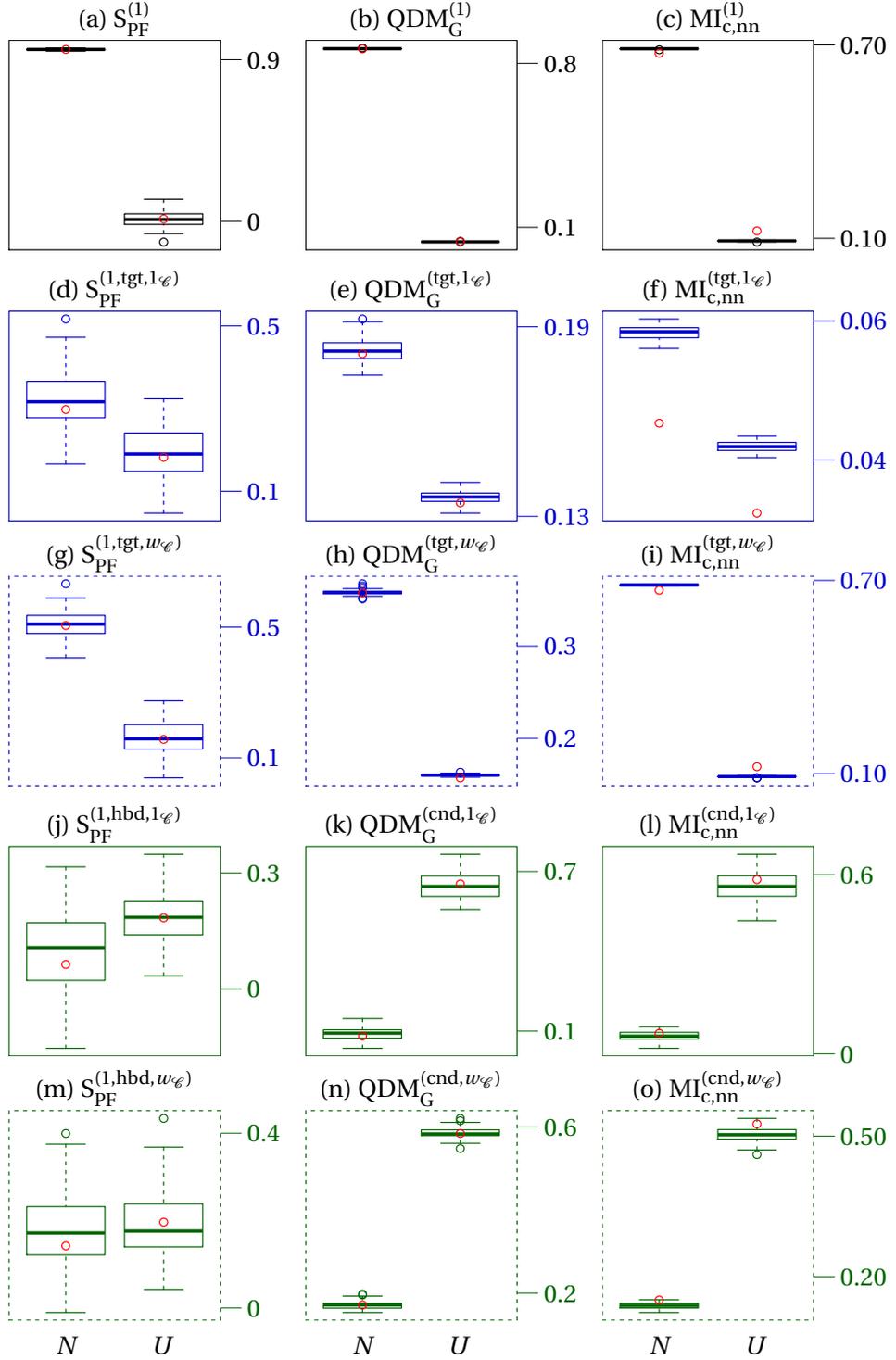

Figure 7: Global (black), target (blue) and conditional (green) sensitivity analysis of minimum normal uniform model on samples of size $n \stackrel{\text{set}}{=} 1\,000$. Red circles are asymptotic values estimated on samples of size $n \stackrel{\text{set}}{=} 10\,000$.

The conditional analysis tells a whole different story: now $U$ is more important than $N$. Indeed, conditionally to both $U$ and $N$ being no less than $c$, $U$ varies in $[c,1]$ while $N$ varies in $[c,+\infty[$, in such a way that the former has more chance to determine the value of their minimum. This is clearly captured by both dependence measures considered, on figures 7(k) and (l), and moreover their smoothed conditional versions improve perceptibly their precision, as indicated by the relative height of the box plots in figures 7(n) and (o), Hybrid correlation ratio adapted to pick-and-freeze estimator follows the same trend on figures 7(j) and (m), but precision is not satisfying at all.

For the Ishigami–Homma model, the critical value is $c = 6.31$. Here, the relative importances of the factors are different in each analysis case. In the global analysis, we have already seen in § 5.1.2 that the factor $X_1$ is the most important, and that the factors $X_2$ and $X_3$ have similar importance, being ranked differently according to different sensibility measures.

In the target analysis, $X_3$ has now similar importance to $X_1$, while $X_2$ has much less. Indeed, the combined effect of $X_1$ and $X_3$ easily exceeds the critical value, while the isolated action of $X_2$ can merely approach the critical value (recall that the parameter $a \stackrel{\text{set}}{=} 5$ is significantly less than $c$). As previously, the dependence measures offer more precision than the correlation ratio with pick-and-freeze estimator. It can be noted that they do not agree exactly on the relative importance of $X_1$ and $X_3$ on figures 8(e) and (f), and that the target quadratic dependence measure does not differ much from its global version in (b). Once again, the smoothed versions are not particularly relevant, and fail completely for the mutual information estimated through copula density.

In the conditional analysis, $X_3$ becomes the dominant factor: being raised to the fourth power, the corresponding term presents steep derivatives in the regime of high values. The mutual information on figure 8(l) seems the most suitable method for putting this into evidence, and once again, the smoothing techniques do improve the quality of both dependence measures considered.

Altogether, these experiments on synthetic data clearly illustrate both the interest of target and conditional sensitivity analysis, and the fact that dependence measures are well suited for this task. Our preliminary results slightly favors the use of mutual information with truncated nearest-neighbors copula density estimation, but more numerical explorations are necessary before drawing further conclusions.

# 6 Conclusion

In the context of sensitivity analysis of complex phenomena in presence of uncertainty, this work motivates and precises the idea of orienting the analysis towards a critical domain of the studied phenomenon. This gives rise to the notions of target and conditional sensitivity analysis. We show that many concepts in the literature relate to them, although usually in more specific frameworks depending on considered applications. Building up on modern statistical tools, we define mathematically a broad range of sensitivity measures which make as few assumptions as possible on the model at hand, while remaining flexible enough to be adapted to many particular situations.

These tools make extensive use of the general concept of nonparametric measure of statistical dependence. We classify a wide variety of available approaches, providing a better understanding of their relationship, while keeping our focus on sensitivity analysis. We extend the theoretical properties of some of them, justifying the expression of estimators which have been proposed previously. We also identify a new way of estimating all Sobol' indices, with a budget of model evaluations which does not depend directly on the number of factors, and without factors independence assumption.

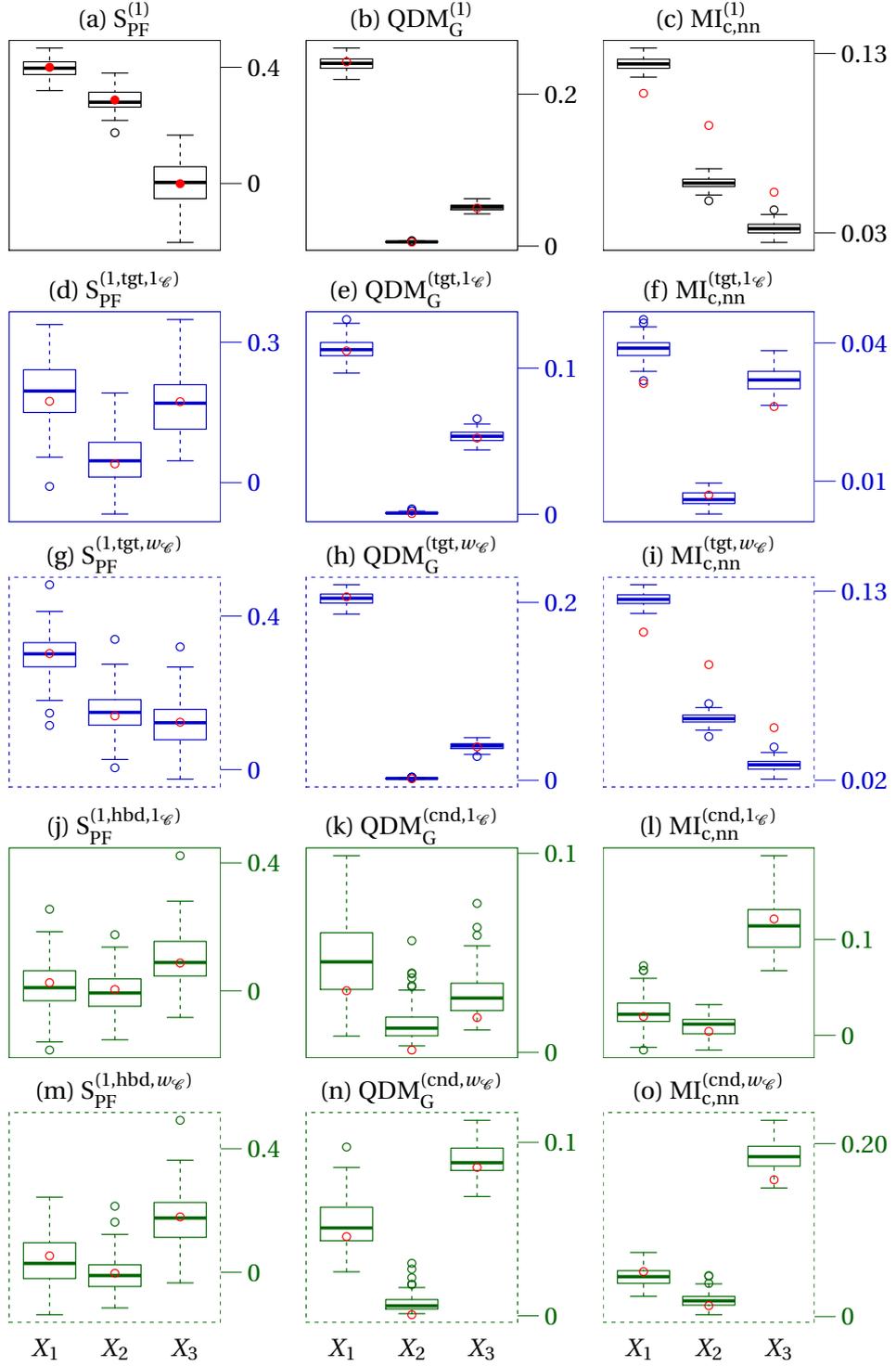

Figure 8: Global (black), target (blue) and conditional (green) sensitivity analysis of Ishigami–Homma model on samples of size $n \stackrel{\text{set}}{=} 1\,000$. Filled red dots are analytical values, hollow red circles are asymptotic values estimated on samples of size $n \stackrel{\text{set}}{=} 10\,000$.

Each and every aspects of this work are illustrated by preliminary numerical experiments, which are encouraging.

Altogether, this work is a good starting point towards sensitivity measures which are more powerful and more adapted to questions raised by experimenters. There is still much to do before actually establishing good practice. First of all, limits and improvements of dependence measures for global sensitivity measures should be further explored, in particular concerning their behavior with multidimensional factors. Then, it seems necessary to analyze theoretically some of the proposed estimators, in particular of the Csiszár divergence dependence measure, and the randomized maximum correlation for the Sobol' indices. Moreover, our tools come in so many different versions (choice of kernels, of functionals, of density estimation, of nonlinear projections) that it would be desirable to compare them in various situations and determine their relative strengths and weaknesses. Then, it is important to test the target and conditional sensitivity measures in more challenging situations, it particular where the critical probability is low, or to put it otherwise, where less critical observations are available. In that respect, we believe that the smoothing technique is promising, if correctly tuned. Last but not least, all these sensitivity measures can only be completely assessed through confrontation to real data.

**Aknowledgemement**

The authors are grateful to Sébastien Da Veiga, SAFRAN, for suggesting the use of dependence measures and for fruitful discussions about target sensitivity analysis.The authors are grateful to Sébastien Da Veiga, SAFRAN, for suggesting the use of dependence measures and for fruitful discussions about target sensitivity analysis.

# A Properties of Csiszár Divergence

We derive here all technical results concerning Csiszár divergences and corresponding dependence measures.

## A.1 Good Properties of Full and Support Csiszár Divergence

Rényi (1959) proposes seven criteria that a dependence measure between random variables should ideally satisfy, and mentions that a certain transformation of the mutual information proposed by Linfoot (1957) satisfies all of them. We do not discuss here Rényi criteria (which are often considered too restrictive), but investigate some properties of the Csiszár divergences that ensure some of them.

The most important to us is the possibility of detecting *any kind* of dependence: the lowest bound is achieved, if, and only if, the variables are independent. In terms of discrepancy between distributions, this corresponds to the *identity of indiscernibles*. We thus consider a measurable space $(\mathcal{Z}, \mathfrak{S})$ and $\mathscr{P}$ the set of all probability distributions over $(\mathcal{Z}, \mathfrak{S})$, and $\phi \colon \mathbb{R}_+ \to \mathbb{R} \cup \{+\infty\}$ a convex function such that $\phi(1) = 0$. We restate first well-known variation range and identity properties of the Csiszár divergences.

**Proposition A.1.** *For all* $P, Q \in \mathscr{P}$, $0 \leq \mathrm{div}_\phi(P, Q) \leq \phi(0) + \phi^*(0)$. *If $\phi$ is strictly convex at* 1, *then for all* $P, Q \in \mathscr{P}$, $\mathrm{div}_\phi(P, Q) = 0$ *is equivalent to* $P = Q$; *if moreover* $\phi(0) + \phi^*(0) < +\infty$, *then for all* $P, Q \in \mathscr{P}$, $\mathrm{div}_\phi(P, Q) = \phi(0) + \phi^*(0)$ *is equivalent to* $P \perp Q$.

A convex function is said *strictly convex* at a point if it is locally linear over *no* neighborhood of this point. We refer the reader to Liese and Vajda (2006) for details and a concise proof of the above. In the following proposition, we shed light on the strict convexity requirement by showing a reciprocal, in particular that it is a necessary condition to identity of indiscernibles.

**Proposition A.2.** *Suppose that $\mathfrak{S}$ is not trivial, that is to say it contains a nonempty strict subset of $\mathcal{Z}$. Then*

(i) $\mathrm{div}_\phi(P, Q) = 0$ *is equivalent to* $P = Q$ *for all* $P, Q \in \mathscr{P}$, *if, and only if, $\phi$ is strictly convex at* 1;

(ii) $\mathrm{div}_\phi(P, Q) = \phi(0) + \phi^*(0)$ *is equivalent to* $P \perp Q$ *for all* $P, Q \in \mathscr{P}$, *if, and only if, $\phi(0) + \phi^*(0)$ is finite and $\phi$ is strictly convex at a point.*

*Proof.* (i). If $\phi$ is not strictly convex at 1, then there exists $\lambda \in \,]0,1[$ and $t_1, t_2 \in \mathbb{R}_+$ such that $t_1 < 1 < t_2$, $\lambda t_1 + (1-\lambda) t_2 = 1$ and $\lambda \phi(t_1) + (1-\lambda) \phi(t_2) = \phi(\lambda t_1 + (1-\lambda) t_2) = \phi(1) = 0$. If now $A \in \mathfrak{S}$ is a nonempty strict subset of $\mathcal{Z}$, let $a \in A$ and $b \in \mathcal{Z} \setminus A$, and define for all $E \in \mathfrak{S}$,

$$P(E) \stackrel{\text{set}}{=} 0 \text{ if } a, b \notin E, \quad \lambda t_1 \text{ if } a \in E, b \notin E, \quad (1-\lambda) t_2 \text{ if } a \notin E, b \in E, \quad 1 \text{ if } a, b \in E, \quad \text{(A.1)}$$

$$Q(E) \stackrel{\text{set}}{=} 0 \text{ if } a, b \notin E, \quad \lambda \text{ if } a \in E, b \notin E, \quad (1-\lambda) \text{ if } a \notin E, b \in E, \quad 1 \text{ if } a, b \in E. \quad \text{(A.2)}$$

It is easy to verify that P and Q are different probability measures over $\mathcal{Z}$, and that Q dominates P with $\frac{dP}{dQ} = t_1 \mathbf{1}_A + t_2 \mathbf{1}_{\mathcal{Z} \setminus A}$, Q-almost everywhere. Thus, $\mathrm{div}_\phi(P, Q) = \lambda \phi(t_1) + (1-\lambda) \phi(t_2) = 0$.
(ii). Supposing that $\phi(0) + \phi^*(0) < +\infty$, we follow Liese and Vajda (2006, proof of Theorem 5) who show that there is no loss of generality in assuming that $\phi$ is nonincreasing over $[0,1]$ and everywhere nonnegative. Thus, both $\phi(0)$ and $\phi^*(0)$ are nonnegative, and $\mathrm{div}_\phi(P, Q) = \phi(0) + \phi^*(0)$ implies $\int \phi\left(\frac{dP_\ll}{dQ}\right) dQ = \phi(0)$ and $\phi^*(0) = \phi^*(0) P_\perp(\mathcal{Z})$. If $\phi$ is strictly convex at $t \in [0,1]$, then by convexity it is strictly decreasing over $[0, t]$, and the former equality implies $\frac{dP_\ll}{dQ} = 0$, Q-almost everywhere. Now if $\phi$ is strictly convex at $t \in [1, +\infty[$, then $\phi^*(0) > 0$ and the last equality implies $P_\perp(\mathcal{Z}) = 1$. In both cases, one gets that $P \perp Q$. Reciprocally, if $\phi(0) = +\infty$, then set for all $E \in \mathfrak{S}$

$$P(E) \stackrel{\text{set}}{=} \quad 0 \text{ if } a, b \notin E, \quad 0 \text{ if } a \in E, b \notin E, \quad 1 \text{ if } a \notin E, b \in E, \quad 1 \text{ if } a, b \in E,$$
$$Q(E) \stackrel{\text{set}}{=} \quad 0 \text{ if } a, b \notin E, \quad \tfrac{1}{2} \in E, b \notin E, \quad \tfrac{1}{2} \text{ if } a \notin E, b \in E, \quad 1 \text{ if } a, b \in E,$$

so that $\text{div}_\phi(P, Q) = +\infty$ but P and Q are not mutually singular. Alternatively, if $\phi^*(0) = +\infty$, exchange the definition of P and Q above to reach the same conclusion. Finally, if $\phi$ is strictly convex at no point then $\phi$ is actually affine, and it is easy to show that for all $P, Q \in \mathscr{P}$, $\text{div}_\phi(P, Q) = \phi(0) + \phi^*(0) = \phi(1) = 0$. ∎

Finally, since we use in our work the support Csiszár divergence, defined by equation 3.4, we would like to know under which conditions similar guarantees are available. As a technical side note, $\mathcal{Z}$ must be a topological space and $\mathfrak{S}$ its Borel $\sigma$-field for the support $\text{supp}(Q) \in \mathfrak{S}$ to be well-defined for any $Q \in \mathscr{P}$. Compare now the previous propositions with the following.

**Proposition A.3.** *Suppose that $\mathfrak{S}$ contains three mutually disjoint nonempty subsets of $\mathcal{Z}$.*

(i) *$\text{sdiv}_\phi(P, Q)$ is nonnegative for all $P, Q \in \mathscr{P}$, if, and only if, $\phi$ is nonnegative over $[0, 1]$.*

(ii) *$\text{sdiv}_\phi(P, Q) = 0$ is equivalent to $P = Q$ for all pair $P, Q \in \mathscr{P}$, if, and only if, either $\phi$ is strictly positive over $[0, 1[$ and strictly convex at 1, or if all finite values of $\phi$ lies in $[0, 1]$ and are strictly negative except at 1.*

(iii) *For all $P, Q \in \mathscr{P}$, $\text{sdiv}_\phi(P, Q) \leq \phi(0) + \max(\phi^*(0), 0)$.*

(iv) *Let $P, Q \in \mathscr{P}$. If $P \perp Q$, then $\text{sdiv}_\phi(P, Q) \geq \phi(0) + \min(\phi^*(0), 0)$. Conversely, suppose that $\phi(0) + \phi^*(0)$ is finite. If $\phi^*(0) \geq 0$ and $\phi$ is strictly convex at a point and $\text{sdiv}_\phi(P, Q) = \phi(0) + \phi^*(0)$, then $P \perp Q$; if moreover $\phi^*(0) > 0$, then $P_\perp(\text{supp}(Q)) = 1$. If $\phi^*(0) < 0$ and $\text{sdiv}_\phi(P, Q) = \phi(0)$, then $P \perp Q$ and $P_\perp(\text{supp}(Q)) = 0$.*

*Proof.* For any $P, Q \in \mathscr{P}$, by Jensen inequality,

$$\text{sdiv}_\phi(P, Q) \geq \phi(P_\ll(\mathcal{Z})) + \phi^*(0) P_\perp(\text{supp}(Q)) ; \tag{A.3}$$

moreover, consider that if $\phi^*(0) < +\infty$, then

$$\text{sdiv}_\phi(P, Q) = \text{div}_\phi(P, Q) - \phi^*(0) P_\perp(\mathcal{Z} \setminus \text{supp}(Q)) . \tag{A.4}$$

(i). Observe that $P_\ll(\mathcal{Z}) \in [0, 1]$, and thus if $\phi$ is nonnegative over $[0, 1]$ and $\phi^*(0) \geq 0$, then $\text{sdiv}_\phi(P, Q) \geq 0$ by equation A.3. Now if $\phi^*(0) < 0$, since $\text{div}_\phi(P, Q) \geq 0$ by proposition A.1, one gets $\text{sdiv}_\phi(P, Q) \geq 0$ with equation A.4. Conversely, suppose that $\phi(t) < 0$ for some $t \in [0, 1[$. If $A \in \mathfrak{S}$ is a nonempty strict subset of $\mathcal{Z}$, let $a \in A$, $b \in \mathcal{Z} \setminus A$, and define for all $E \in \mathfrak{S}$,

$$Q(E) \stackrel{\text{set}}{=} \begin{cases} 0 & \text{if } a \notin E, \\ 1 & \text{if } a \in E, \end{cases} \quad P_\ll(E) = \begin{cases} 0 & \text{if } a \notin E, \\ t & \text{if } a \in E, \end{cases} \quad P_\perp(E) = \begin{cases} 0 & \text{if } b \notin E, \\ 1-t & \text{if } b \in E, \end{cases} \tag{A.5}$$

and $P \stackrel{\text{set}}{=} P_\ll + P_\perp$. It is easy to verify that $P, Q \in \mathscr{P}$, $P_\perp \perp Q$ and $P_\ll \ll Q$, with $\frac{dP_\ll}{dQ} = t1_A$, Q-almost everywhere. Finally, $\text{sdiv}_\phi(P, Q) = \phi(t) Q(A) = \phi(t) < 0$.

(ii). If $P = Q$, then obviously $\text{sdiv}_\phi(P, Q) = 0$. Suppose now that $\text{sdiv}_\phi(P, Q) = 0$, and that $\phi$ is strictly positive over $[0, 1[$ and strictly convex at 1. If $\phi^*(0) \leq 0$, then by Jensen inequality equation A.3 shows that $\phi(P_\ll(\mathcal{Z})) = 0$ and thus $P_\ll(\mathcal{Z}) = 1$, so that $P = P_\ll$, hence $P \ll Q$. But in such case, $\text{div}_\phi(P,Q) = \text{sdiv}_\phi(P,Q) = 0$. If $\phi^*(0) < 0$, it is now (i) together with equation A.4 which shows that $\text{div}_\phi(P,Q) = 0$. In both cases, this implies $P = Q$ by proposition A.1. Now if $\phi$ takes negative values in $[0,1[$ but no finite strictly positive values, then $\phi^*(0) = +\infty$, and $\text{sdiv}_\phi(P,Q) = 0$ implies $P_\perp(\text{supp}(Q)) = 0$ and $\int \phi\left(\frac{dP_\ll}{dQ}\right) dQ = 0$ and thus $\frac{dP_\ll}{dQ} = 1$, Q-almost everywhere; that is, $P = Q$.

Conversely, suppose first that $\phi$ is not strictly convex at 1; then P, Q as defined by equations A.1 and A.2 in proposition A.2 are different and satisfy $\text{sdiv}_\phi(P,Q) = 0$. Suppose otherwise that $\phi$ vanishes at some $t \in [0,1[$; then P, Q as defined by equation A.5 are different and satisfy $\text{sdiv}_\phi(P,Q) = \phi(t) = 0$. Suppose finally that $\phi(t_1) < 0$ for some $t_1 \in [0,1[$ and $0 < \phi(t_2) < +\infty$ for some $t_2 \in \mathbb{R}_+$. The former implies by convexity that $\phi$ is strictly positive over $]1,+\infty[$, and with the latter we can define $\lambda \stackrel{\text{def}}{=} \frac{\phi(t_2)-\phi(t_1)}{\phi(t_2)} \in ]0,1[$ so that $\lambda\phi(t_1) + (1-\lambda)\phi(t_2) = 0$. Again by convexity, $\phi(\lambda t_1 + (1-\lambda)t_2) \leq 0$, and necessarily $\lambda t_1 + (1-\lambda) t_2 \leq 1$. Let then $A, B, C \in \mathfrak{S}$ be nonempty disjoint subsets of $\mathcal{Z}$, let $a \in A$, $b \in B$ and $c \in C$, and define for all $E \in \mathfrak{S}$,

$$P(E) \stackrel{\text{set}}{=} \sum_{z \in \{a,b,c\}} p_z 1_E(z) \quad \text{where} \quad p_a = \lambda t_1, \quad p_b = (1-\lambda)t_2, \quad \text{and} \quad p_c = 1 - p_a - p_b,$$

and $Q(E)$ just as in equation A.2. Once again, it is easy to verify that P and Q are different probability measures over $\mathcal{Z}$, and that $\text{div}_\phi(P,Q) = \lambda\phi(t_1) + (1-\lambda)\phi(t_2) = 0$.

(iii). If $\phi^*(0) = +\infty$ the inequality is trivial; otherwise, it follows from the upper bound of proposition A.1 and from equation A.4 that $\text{sdiv}_\phi(P,Q) \leq \phi(0) + \phi^*(0)P_\perp(\text{supp}(Q))$, and the result is deduced considering that $P_\perp(\text{supp}(Q)) \in [0,1]$.

(iv). If $P \perp Q$, then $P_\ll = 0$ and thus $\text{sdiv}_\phi(P,Q) = \phi(0) + \phi^*(0)P_\perp(\text{supp}(Q))$ and the result follows as above. Conversely, suppose that the upper bound in (iii) is reached. If $\phi^*(0) > 0$, then necessarily $P_\perp(\text{supp}(Q)) = 1$, hence $\text{sdiv}_\phi(P,Q) = \text{div}_\phi(P,Q)$. If $\phi^*(0) = 0$, the same equality is straightforward. In both cases, one obtains $P \perp Q$ from proposition A.1 under strict convexity condition. Finally, if $\phi^*(0) < 0$, then by convexity $\phi$ is strictly decreasing, and necessarily $P_\perp(\text{supp}(Q)) = 0$ and $\frac{dP_\ll}{dQ} = 0$, Q-almost everywhere, that is $P \perp Q$. ∎

**Remark A.1.** Proposition A.3 (i) and (ii) essentially say that good properties of Csiszár divergences are kept for the support Csiszár divergences, provided that $\phi$ is positive over $[0,1]$. In (ii), the case with no finite positive values is somewhat pathological and not useful in practice.

**Remark A.2.** Unfortunately, proposition A.3 (iv) cannot be written as an equivalence (except when $\phi^*(0) = 0$, in which case $\text{sdiv}_\phi = \text{div}_\phi$). Indeed, the lower bound $\text{sdiv}_\phi(P,Q) \geq \phi(0) + \min(\phi^*(0), 0)$ for mutually singular P, Q cannot be improved, since one can always construct instances of equality. This means that our support Csiszár divergence does not always identify cases of mutual singularity. This is however a mild limitation for the purpose of sensitivity analysis through dependence measure, where the most crucial need is to rule out cases of independence; this ability is ensured by proposition A.3 (ii).

### A.2 Weighted Csiszár Divergence

Weighted Csiszár divergence is introduced in § 4.4.3. In the same setting as above, let moreover $w: \mathcal{Z} \to \mathbb{R}_+$ be a measurable function such that $0 < \int w \, dP < +\infty$ and $0 < \int w \, dQ < +\infty$. First, let us study Lebesgues decompositions of probabilities weighted by the same weights.

**Proposition A.4.** *The Lebesgues decomposition of $P^w$ with respect to $Q^w$ is $P^w = P^w_\ll + P^w_\perp$, where*

$$P^w_\ll: A \mapsto \left(\int w \, dP\right)^{-1} \int_A w \, dP_\ll \quad \text{and} \quad P^w_\perp: A \mapsto \left(\int w \, dP\right)^{-1} \int_A w \, dP_\perp$$

*are the absolutely continuous and singular parts, respectively. Moreover, it holds that*

$$\frac{dP^w_{\ll}}{dQ^w} = \frac{\int w\,dQ}{\int w\,dP} \times \frac{dP_{\ll}}{dQ}, \qquad Q^w\text{-almost surely.}$$

*Proof.* Let $A \in \mathfrak{S}$. Then by definition, $P^w(A) = \left(\int w\,dP\right)^{-1} \int_A w\,dP = \left(\int w\,dP\right)^{-1} \int_A w\,dP_{\ll} + \left(\int w\,dP\right)^{-1} \int_A w\,dP_{\perp}$. First, note that if $N \in \mathfrak{S}$ is such that $Q(N) = P_{\perp}(\mathcal{Z} \setminus N) = 0$, then obviously $\int_N w\,dQ = \int_{\mathcal{Z}\setminus N} w\,dP_{\perp} = 0$, so that $P^w_{\perp}$ is singular with respect to $Q^w$. Now by absolute continuity, $\int_A w\,dP_{\ll} = \int_A w\frac{dP_{\ll}}{dQ}\,dQ = \left(\int w\,dQ\right)\int_A \frac{dP_{\ll}}{dQ}\,dQ^w$, hence $Q^w = 0$ implies $P^w_{\ll}(A) = 0$; the conclusion follows from uniqueness of the Radon–Nikodym derivative. ∎

The general expression of Csiszár divergence between $P^w$ and $Q^w$ can be deduced easily; the following explicits the particular case where $w$ has equal total mass under P and Q, fully justifying the name *weighted Csiszár divergence*.

**Corollary A.1.** *If $\int w\,dP = \int w\,dQ$, then*

$$\mathrm{div}^w_\phi(P,Q) = \left(\int w\,dP\right)^{-1}\left(\int \phi\left(\frac{dP_{\ll}}{dQ}\right)w\,dQ + \phi^*(0)\int w\,dP_{\perp}\right),$$

*and*

$$\mathrm{sdiv}^w_\phi(P,Q) = \left(\int w\,dP\right)^{-1}\left(\int \phi\left(\frac{dP_{\ll}}{dQ}\right)w\,dQ + \phi^*(0)\int 1_{\mathrm{supp}(Q^w)} w\,dP_{\perp}\right).$$

*Proof.* By definition of weighted probabilities together with proposition A.4, $\mathrm{div}^w_\phi(P,Q) = \int \phi\left(\frac{dP^w_{\ll}}{dQ^w}\right)dQ^w + \phi^*(0)P^w_{\perp}(\mathcal{Z}) = \frac{1}{\int w\,dQ}\int \phi\left(\frac{\int w\,dQ}{\int w\,dP}\frac{dP_{\ll}}{dQ}\right)w\,dQ + \frac{\phi^*(0)}{\int w\,dP}\int w P_{\perp}$, leading to the result with the additional hypothesis of equal total masses. ∎

**Remark A.3.** Strictly speaking, one cannot simply replace $\mathrm{supp}(Q^w)$ by $\mathrm{supp}(Q)$ in the support version, because there might be points in the latter and not in the former, associated with non zero weights. However, such points precisely have a neighborhood with Q-almost zero weights so that they have no influence in an estimator of the form $\int_{\mathrm{supp}(Q)}\phi\left(\frac{dP_n}{dQ_n}\right)w\,dQ$, where $\frac{dP_n}{dQ_n}$ is any finite estimation of the Radon–Nikodym derivatives. There is thus no need for estimating $\mathrm{supp}(Q^w)$.

Finally the last propositions give properties concerning only weighted probabilities, for image measures and product measures; they are useful for characterizing weighted Csiszár divergence dependence measures.

**Proposition A.5.** *Let $(\Omega, \mathfrak{F}, P)$ be a probability space, $X$ be a random variable, and $w \colon \mathrm{ran}(X) \mapsto \mathbb{R}_+$ be measurable such that $\mathrm{E}(w(X))$ is nonzero and finite. Then, $\left(P^{w(X)}\right)_X = \left(P_X\right)^w$.*

*Proof.* Let $A \subset \mathrm{ran}(A)$ be measurable. In the one hand, $\left(P^{w(X)}\right)_X(A) = P^{w(X)}(X^{-1}(A)) = \mathrm{E}(1_A(X)w(X))/\mathrm{E}(w(X))$. In the other hand, $\left(P_X\right)^w(A) = \int 1_A w\,dP_X / \int w\,dP_X$. Integrating against the image measure yields the equality. ∎

**Proposition A.6.** *Let $P$, $Q$ be probability distributions over sets $\mathcal{X}$ and $\mathcal{Y}$, respectively, and $w_{\mathcal{X}} \colon \mathcal{X} \to \mathbb{R}_+$, $w_{\mathcal{Y}} \colon \mathcal{Y} \to \mathbb{R}_+$ be measurable functions such that $\int w_{\mathcal{X}}\,dP$ and $\int w_{\mathcal{Y}}\,dQ$ are nonzero and finite. Then, defining $w_{\mathcal{X}\times\mathcal{Y}} \colon \mathcal{X} \times \mathcal{Y} \to \mathbb{R}_+ \colon (x,y) \mapsto w_{\mathcal{X}}(x)w_{\mathcal{Y}}(y)$, the weighed probability $(P \otimes Q)^{w_{\mathcal{X}\times\mathcal{Y}}}$ is well-defined and equal to $P^{w_{\mathcal{X}}} \otimes Q^{w_{\mathcal{Y}}}$.*

*Proof.* Let $A \subset \mathcal{X}$ and $B \subset \mathcal{Y}$ be measurable sets. Then by separability $\int_{A\times B} w_{\mathcal{X}\times\mathcal{Y}}\,d(P\otimes Q) = \int_A w_{\mathcal{X}}\,dP \int_B w_{\mathcal{Y}}\,dQ$. The particular case $A \stackrel{\text{set}}{=} \mathcal{X}$ and $B \stackrel{\text{set}}{=} \mathcal{Y}$ shows that $(P \otimes Q)^{w_{\mathcal{X}\times\mathcal{Y}}}$ is well-defined; and the general case that $(P\otimes Q)^{w_{\mathcal{X}\times\mathcal{Y}}}(A\times B) = \frac{\int_A w_{\mathcal{X}}\,dP}{\int w_{\mathcal{X}}\,dP}\frac{\int_B w_{\mathcal{Y}}\,dQ}{\int w_{\mathcal{Y}}\,dQ} = P^{w_{\mathcal{X}}}(A)Q^{w_{\mathcal{X}}}(B)$. The conclusion follows, since the product measure is characterized over the rectangles. ∎